\def\webDOI{http://dx.doi.org}

\documentclass[final,2p,times]{amsart}

\usepackage{epsfig}
\usepackage{subfigure}
\usepackage{amsmath}
\usepackage{amsfonts}
\usepackage{amssymb}
\usepackage{mathrsfs}
\usepackage{bbm}
\usepackage{slashbox}
\usepackage{cite}
\usepackage[pdftex, pdfborderstyle={/S/U/W 1}]{hyperref}
\usepackage{nccmath}

 \newcommand{\eqnref}[1]{(\ref{#1})}

\newcommand{\ci}{\mathrm{i}}
\newcommand{\Nset}{\mathbb{N}}
\newcommand{\Rset}{\mathbb{R}}

\newcommand{\Pset}{\mathbb{P}}
\newcommand{\itr}{{\sf T}}
\newcommand{\xgj}{x}
\newcommand{\ygj}{y}
\newcommand{\xigj}{\xi}
\newcommand{\xg}{{\boldsymbol\xgj}}
\newcommand{\yg}{{\boldsymbol\ygj}}
\newcommand{\xig}{{\boldsymbol\xigj}}
\newcommand{\Xig}{{\boldsymbol\Xi}}
\newcommand{\QoI}{\ygj}
\newcommand{\QOI}{Y}
\newcommand{\vQoI}{\yg}
\newcommand{\QoIset}{{\mathcal Y}}
\newcommand{\vj}{{\boldsymbol j}}
\newcommand{\model}{g}
\newcommand{\modelg}{{\boldsymbol\model}}
\newcommand{\hgj}{h}
\newcommand{\hg}{{\boldsymbol\hgj}}
\newcommand{\vbase}{\psi}
\newcommand{\mbase}{{\boldsymbol\Psi}}
\newcommand{\vsensing}{\varphi}
\newcommand{\msensing}{{\boldsymbol\Phi}}
\newcommand{\weight}{\omega}
\newcommand{\mweight}{{\boldsymbol W}}
\newcommand{\vmeasure}{{\boldsymbol m}}
\newcommand{\mmeasure}{{\boldsymbol M}}
\newcommand{\cubature}{{\boldsymbol\Theta}}
\newcommand{\basis}{{\mathcal B}}
\newcommand{\coefj}{c}
\newcommand{\coefv}{{\boldsymbol\coefj}}
\newcommand{\nquad}{n}
\newcommand{\torder}{p}
\newcommand{\level}{N}
\newcommand{\cardinal}{P}
\newcommand{\heps}{\varepsilon}

\newcommand{\demi}{\frac{1}{2}}

\newcommand{\iexp}{\mathrm{e}}
\newcommand{\id}{\mathrm{d}}
\newcommand{\eref}[1]{Eq.~\eqnref{#1}}
\newcommand{\fref}[1]{Fig.~(\ref{#1})}
\newcommand{\nominal}[1]{\underline{#1}}
\newcommand{\surr}[1]{\hat{#1}}
\newcommand{\esp}[1]{\mathbb{E}\{\smash{#1}\}}
\newcommand{\dual}[1]{\langle\smash{#1}\rangle}
\newcommand{\cjg}[1]{{#1}^*}

\newtheorem{mytheorem}{Theorem}[section]

\begin{document}

\title[Sparse polynomial surrogates in CFD]{Sparse polynomial surrogates for aerodynamic computations with random inputs}

\author[\'E. Savin]{\'Eric Savin}
\address[\'E. Savin]{Onera--The French Aerospace Lab, France}
\thanks{Corresponding author: \'E. Savin, Onera--The French Aerospace Lab, Computational Fluid Dynamics Dept., 29, avenue de la Division Leclerc, F-92322 Ch\^atillon cedex, France (Eric.Savin@onera.fr).}
\email{eric.savin@onera.fr}

\author[A. Resmini]{Andrea Resmini}
\address[A. Resmini]{Pierre-et-Marie-Curie University -- d'Alembert Institute \& Onera--The French Aerospace Lab, France}
\email{andrea.resmini@onera.fr}

\author[J. Peter]{Jacques Peter}
\address[J.Peter]{Onera--The French Aerospace Lab, France}
\email{jacques.peter@onera.fr}

\begin{abstract}
This paper deals with some of the methodologies used to construct polynomial surrogate models based on generalized polynomial chaos (gPC) expansions for applications to uncertainty quantification (UQ) in aerodynamic computations. A core ingredient in gPC expansions is the choice of a dedicated sampling strategy, so as to define the most significant scenarios to be considered for the construction of such metamodels. A desirable feature of the proposed rules shall be their ability to handle several random inputs simultaneously. Methods to identify the relative "importance" of those variables or uncertain data shall be ideally considered as well. The present work is more particularly dedicated to the development of sampling strategies based on sparsity principles. Sparse multi-dimensional cubature rules based on general one-dimensional Gauss-Jacobi-type quadratures are first addressed. These sets are non nested, but they are well adapted to the probability density functions with compact support for the random inputs considered in this study. On the other hand, observing that the aerodynamic quantities of interest (outputs) depend only weakly on the cross-interactions between the variable inputs, it is argued that only low-order polynomials shall significantly contribute to their surrogates. This "sparsity-of-effects" trend prompts the use of reconstruction techniques benefiting from the sparsity of the outputs, such as compressed sensing (CS). CS relies on the observation that one only needs a number of samples proportional to the compressed size of the outputs, rather than their uncompressed size, to construct reliable surrogate models. The results obtained with the test case considered in this work corroborate this expected feature.
\end{abstract}



\maketitle

\section*{Nomenclature}

\begin{tabbing}
  XXXXX \= \kill
  $\basis$ \> Polynomial basis \\
  $c$ \> Chord length\\
  $C_D$ \> Drag coefficient\\
  $C_L$ \> Lift coefficient\\
  $C_M$ \> Pitching moment coefficient\\
  $C_p$ \> Static pressure coefficient\\
  $C_{p1}^*$ \> Total pressure coefficient\\
  $D$ \> Parameters space dimension\\
  $\esp{\cdot}$ \> Mathematical expectation\\
  $\model$ \>  Generic physical model\\
  $\surr{\model}$ \>  Surrogate model\\
  $\modelg$ \>  Discretized computational model\\
  $\model_j$ \> Polynomial chaos expansion coefficient of the computational model\\
  $h$ \> Thickness\\
  $M_\infty$ \> Free-stream Mach number\\
  $\mmeasure$ \> Measurement matrix\\
  $\nquad$ \> Number of cubature points\\
  $\level$ \> Level of the cubature rule\\ 
  $\torder$ \> Total order of the polynomial surrogates\\
  $\cardinal$ \> Number of polynomials\\
  $P_\Xig$ \> Probability density function of the random input parameters\\
  $\Pset^\torder[\xg]$ \> Polynomial set of total order $\torder$\\
  $r$ \> Thickness-to-chord ratio\\
  $Re$ \> Reynolds number\\
  $S$ \> Sparsity (the number of non-zero entries)\\
  $\vQoI$ \> Output quantities of interest\\
  $\mweight$ \>Weighting matrix\\
  $\alpha$ \> Angle of attack\\
  $\beta_{\mathrm I}$ \> Beta law of the first kind\\
  $\beta_2$ \> Kurtosis\\
  $\delta_S$ \> Restricted isometry constant\\
  $\heps$ \> Polynomial truncation error\\
  $\gamma_1$ \> Skewness\\
  $\mu$ \> Mean\\
  $\mu(\msensing,\mbase)$ \> Mutual coherence\\
  $\vsensing_\ell$ \> $\ell$-th sensing vector\\
  $\msensing$ \> Sensing basis\\
  $\vbase_j$ \> $j$-th representation vector (polynomial chaos)\\
  $\mbase$ \> Representation basis \\
  $\sigma$ \> Standard deviation\\
  $\Theta_1$ \> One-dimensional quadrature rule\\
  $\cubature$ \> Cubature set\\
  $\xig$ \> Random input parameters\\
  $\weight_\ell$ \> $\ell$-th cubature weight\\
  $\xig_\ell$ \> $\ell$-th cubature point in the parameters space\\[5pt]
  \textit{Subscript}\\
  $j$ \> Index of the polynomial chaos\\
  $\ell$ \> Index of a cubature point\\
 \end{tabbing}

\section{Introduction}

Surrogate models are typically used to perform optimization or uncertainty quantification (UQ) studies involving a complex modeling and simulation process, as encountered in computational fluid dynamics (CFD) among other engineering sciences applications. The principle of a surrogate model relies on an interpolation or regression procedure to estimate a scalar or vector field using a sampling data set constituted by carefully chosen outputs of the aforementioned complex process. Since the latter is often computationally expensive to run, the surrogate model shall be able to work out reliable estimates of its outputs at no extra computational costs except the evaluation of the surrogate itself. It should thus offer a non intrusive alternative to expensive runs of a complex process in order to sweep across the parameter space that influences it. When considering large parameter spaces, efficient algorithms are needed to provide an accurate surrogate representation of such parametric outputs. The kriging procedure~\cite{krige:1951} has gained a large attention over the past decades due to its robustness, accuracy, and ability to provide an estimate of the error done by the procedure; see the reviews by Kleijnen~\cite{kleijnen:2009} or Bompard~\cite{BOM11} and references therein. It has been applied to the simulation of air flow around a wing profile with the consideration of variable geometrical parameters in a previous study~\cite{Onera14}. 

The polynomial chaos (PC) expansion is also a powerful tool for constructing spectral-like surrogate models of a parameterized process. A general methodology based on the Galerkin method has originally been proposed by Sun~\cite{SUN79} and Ghanem \& Spanos~\cite{Ghanem:2003} for the computation of the PC coefficients of the solution of a parameterized partial differential equation (PDE). This original scheme is heavily intrusive and has prompted the development of non-intrusive schemes, especially for PDEs arising in fluid dynamic models~\cite{OLM:2010}. Two approaches for computing the coefficients of a PC expansion have usually been considered: (i) a projection approach, in which they are computed by structured (\emph{i.e.} Gauss quadratures) or unstructured (\emph{i.e.} Monte-Carlo or quasi Monte-Carlo sampling) quadratures; or (ii) a regression approach, minimizing some error measure or truncation tolerance. Both techniques suffer from some well identified shortcomings when the dimension of the parameter space, and the number of model evaluations alike, increase. Indeed, a PC expansion of total degree $p$ in $D$ variable parameters contains $\smash{\frac{(p+D)!}{p!D!}}$ coefficients. A direct way to compute them is to use a tensor product grid in the parameter space requiring $\level^D$ evaluations of the process, where the number $\level$ of evaluations needed for one particular direction in that space is related to $p$. These $N^D$ runs are very often unaffordable for large parameter spaces and complex configurations, as in CFD for example. Fortunately, the Smolyak algorithm~\cite{SMO63} defines sparse grid quadratures involving ${\mathrm O}(\level^{\log D})$ points while preserving a satisfactory level of accuracy. Consequently, collocation techniques with sparse quadratures or adaptive regression strategies have been developed in order to circumvent this dimensionality concern~\cite{ELD09,BLA10,GHI02,XIU05}. 

Other directions may be considered to deal with large parameter spaces. In the work presented in this paper we aim at benefiting from the sparsity of the process outputs themselves to reconstruct their PC representations in a non-adaptive way \cite{DOO11,MAT12}. Indeed, we rely on the common observation that many cross-interactions between the input parameters are actually smoothened, or even negligible, once that have been propagated to some global quantities of interest processed from, say, complex aerodynamic computations. We can therefore expect to achieve a successful output recovery by the techniques known under the terminology of compressed sensing (CS)~\cite{CAN06,DON06}. In this theory the reconstruction of a sparse signal on a given, known basis requires only a limited number of evaluations at randomly selected points--at least significantly less than the \emph{a priori} dimension of the basis.  We thus resort to unstructured sampling sets to recover sparse outputs. We may also resort to highly structured sampling sets, such as nesting sets in large parameter spaces. The objective in this case is to be able to enrich the surrogate models by structured samples compatible with the structured samples used in a previous evaluation, without the need to re-evaluate the whole process. Classical nested quadratures in one dimension go from a set of $\level$ points to a set of $2\level\pm1$ points (for Fej\'er second rule~\cite{Fej_33}) or $2\level-1$ points (for Clenshaw-Curtis rule \cite{CleCur_60,Tre_08}), such that nested rules involve sets of nodes of which dimensions $\nquad$ double each time their so-called level $\level$ is incremented, $\nquad\sim2^\level$. These sets can subsequently by used with the Smolyak algorithm~\cite{SMO63} to construct nested sparse sets in higher dimensions, together with adaptive strategies as developed by Gerstner \& Griebel~\cite{GER98} for example.

We will not consider nested rules in this paper though, leaving these developments to another publication. As invoked above, we rather focus on sparse multi-dimensional cubature rules based on one-dimensional Gauss-type quadratures, on one hand. These sets are non nested, but they are well adapted to probability density functions with compact supports for the input parameters considered in the example addressed in this work. On the other hand, the "sparsity-of-effects" trend observed in complex simulations prompts the use of dedicated reconstruction techniques benefiting from the sparsity of the outputs. Since only low-order polynomials will significantly contribute to the output surrogates, one may infer from CS theory that only a number of samples proportional to the compressed size of the output, rather than the uncompressed size, is actually needed. In addition, these sampling points should be chosen randomly--\emph{i.e.} they are unstructured. Therefore we may invoke sparsity patterns either for the input parameters or the output quantities of interest, to construct polynomial surrogates of the latter.

The rest of the paper is organized as follows. Section \ref{sec:BC02} introduces the basic configuration and CFD tools, namely a two-dimensional RAE 2822 transonic airfoil~\cite{COO79,nasa} and the Onera in-house \emph{elsA} software\cite{CAM13,elsa}. This example serves as a guideline for the metamodeling techniques based on generalized PC expansions introduced in the subsequent sections: construction by quadrature rules (structured grids in the parameter space) in section \ref{sec:gPC} or by compressed sensing (using unstructured grids) in section \ref{sec:CS}, while the intermediate section \ref{sec:gPC-BC02} details the particular application of the former approach to the RAE 2822 airfoil. Some general conclusions and perspectives are drawn in section \ref{sec:conclusions}.

\section{Problem setup}\label{sec:BC02}

We start by introducing the problem setup and the numerical tools used to solve it. We consider a two-dimensional transonic flow around an RAE 2822 transonic airfoil solved by steady-state Reynolds-Averaged Navier-Stokes (RANS) equations together with a Spalart-Allmaras turbulence model closure~\cite{SPA92}. The nominal flow conditions are the ones described in Cook \emph{et al.}~\cite{COO79} for the test case \#6 together with the wall interference correction formulas derived in~\cite[pp.~386--387]{GAR66} and their slight modifications suggested in~\cite[p.~130]{HAA93} (see also the CFD verification and validation web-site of the NPARC Alliance \cite{nasa}). The nominal free-stream Mach number $\nominal{M}_\infty=0.729$, angle of attack $\nominal{\alpha}=2.31^\text{o}$, and Reynolds number $\nominal{Re}=6.50\cdot10^6$ (based on the airfoil chord length $c$, fluid velocity, temperature and molecular viscosity at infinity) arise from the corrections $\Delta M_\infty=0.004$ and $\Delta\alpha=-0.61^\text{o}$ given in~\cite[p.~130]{HAA93} for the test case \#6 outlined in Cook \emph{et al.}~\cite{COO79}, for which $\nominal{M}_\infty=0.725$,  $\nominal{\alpha}=2.92^\text{o}$, and $\nominal{Re}=6.50\cdot10^6 $.

\subsection{Numerical nominal model}

\begin{figure}[t]
\centering{\includegraphics[width=7.4cm]{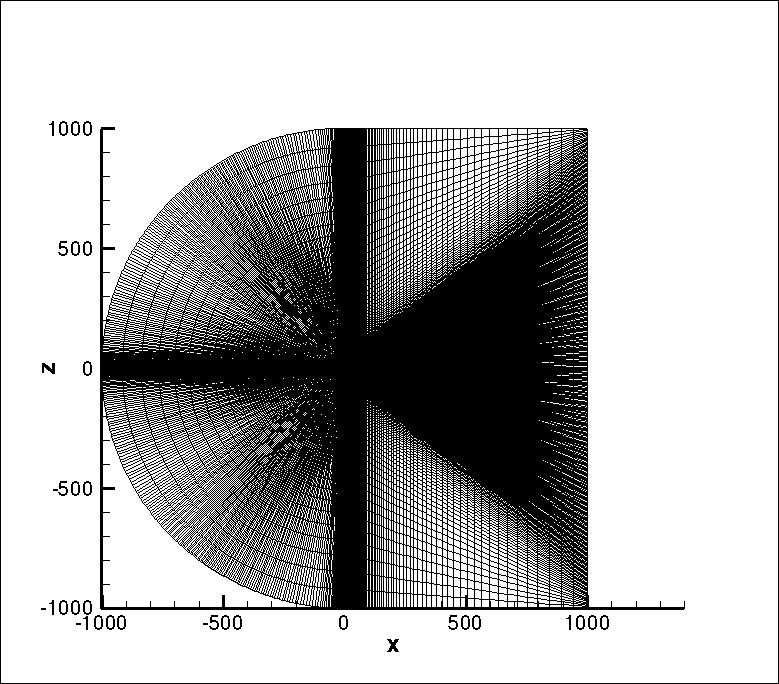}}
\caption{Computational domain for the baseline configuration.}\label{fg:rae_mesh}
\end{figure}

\begin{figure}[t]
\centering{\includegraphics[width=7.4cm]{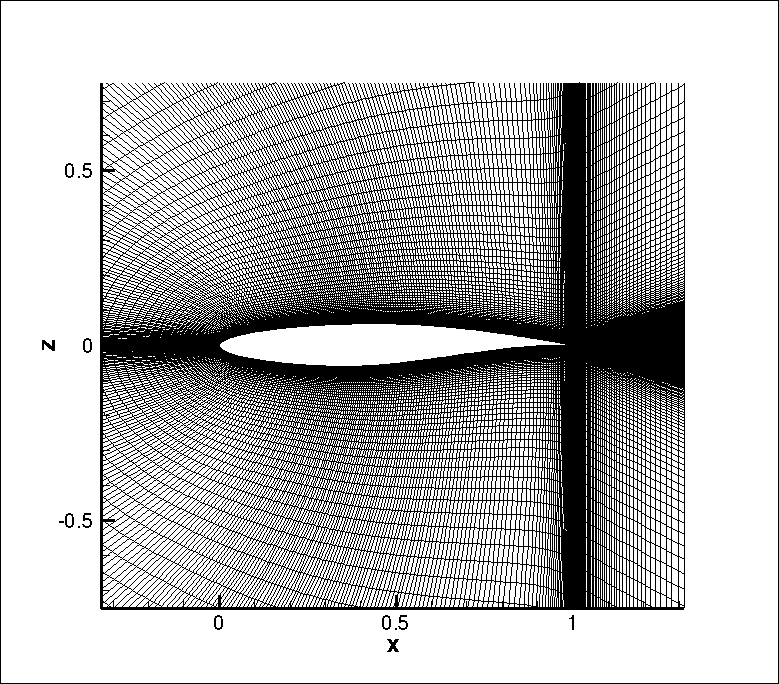}}
\caption{Close view of the mesh around the RAE 2822 aerofoil for the baseline configuration.}\label{fg:rae_mesh_close}
\end{figure}

\begin{figure}[t]
\centering{\includegraphics[width=7.4cm]{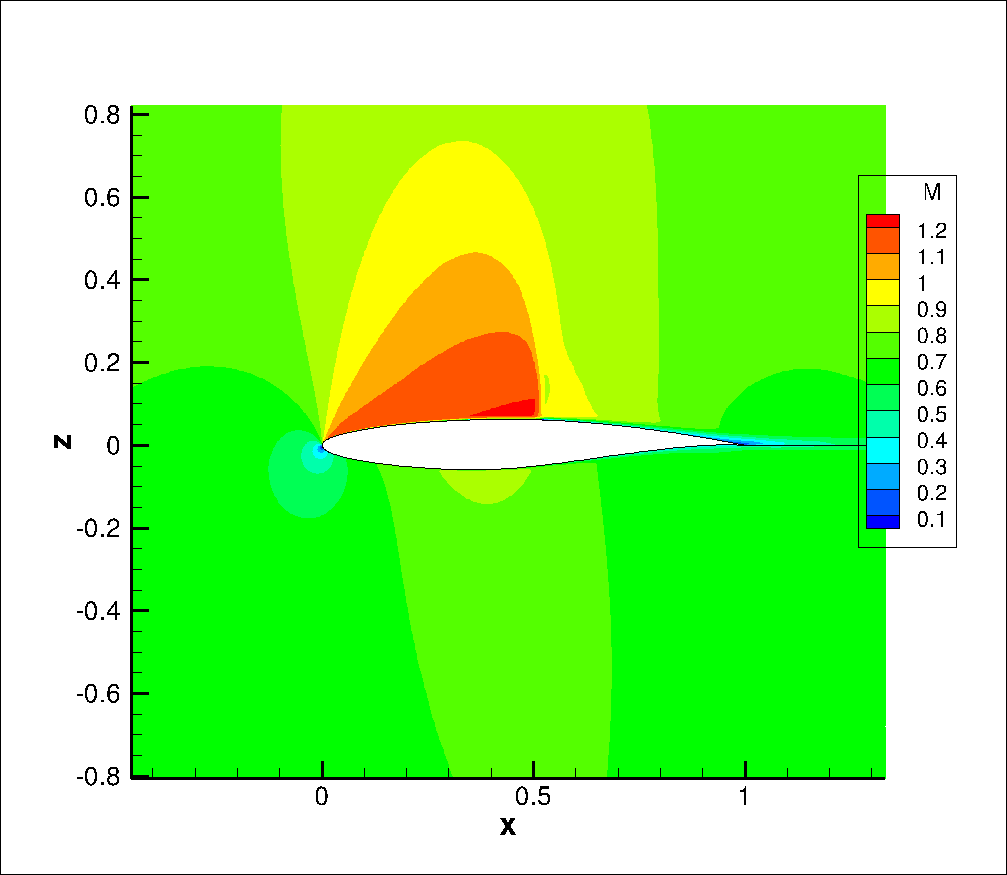}}
\caption{Magnitude of velocity for the baseline RAE 2822 transonic airfoil at $M_\infty=0.729$, $\alpha=2.31^\text{o}$, $Re=6.50\cdot10^6$.}\label{fig:rae_flow_Mach}
\end{figure}

\begin{figure}[t]
\centering{\includegraphics[width=8.4cm]{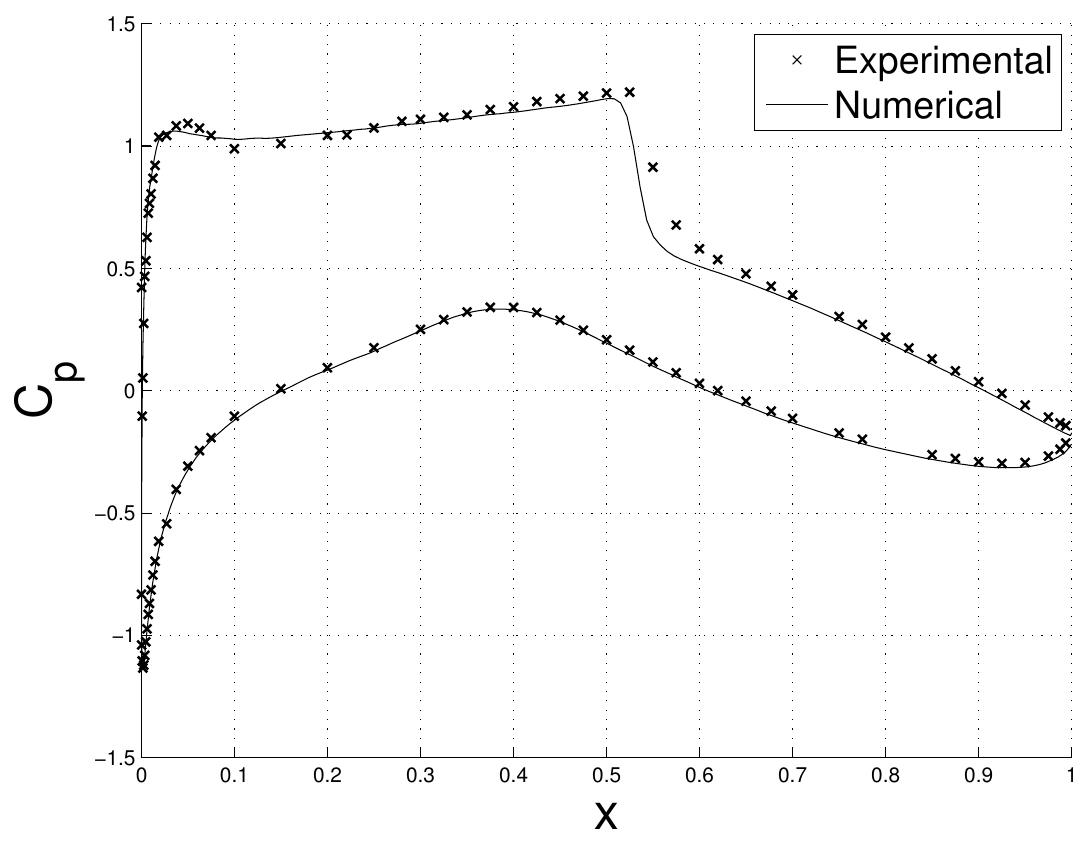}}
\caption{Static pressure coefficient $C_p$ at the wall for the baseline RAE 2822 transonic airfoil at $M_\infty=0.729$, $\alpha=2.31^\text{o}$, $Re=6.50\cdot10^6$, compared with the experiment results gathered on the CFD verification and validation web-site of the NPARC Alliance~\cite{nasa} (crosses).} \label{fig:rae_flow_Cp}
\end{figure}

The nominal problem is discretized using a $769c\times193c$ mesh shown in \fref{fg:rae_mesh} and \fref{fg:rae_mesh_close}, where the boundary at infinity was left intensionally far (at about $500c$ from the airfoil). These values proved to be sufficient to avoid spurious reflection with the far-field boundary. The discretized numerical solution is computed using the cell-centered finite volume CFD software \emph{elsA} \cite{CAM13,elsa}. It is based on:
\begin{itemize}
\item Roe flux using a second order MUSCL scheme~\cite{VLE79} (based on van Albada limiter~\cite{VAL82}) for the convective term of the RANS system;
\item First order Roe fluxes for the advection term of the turbulent variable;
\item Corrected second order diffusive terms based on the corrected mean of the cell-centered gradients of the two adjacent cells (referred to as the "5p\_cor" approach);
\item Source terms for the turbulent transport computed using the temperature gradients at the center of the cells.
\end{itemize}
The flow is attached with a weak shockwave on the suction side. The contour plot of the magnitude of the velocity are displayed on \fref{fig:rae_flow_Mach} and the static pressure profile at the wall are displayed on \fref{fig:rae_flow_Cp}. Given the large number of simulations to run, the numerical parameters of the steady state algorithm proved to be essential to insure a fast convergence. This was performed using the following:
\begin{itemize}
\item An implicite algorithm based on the Lower-Upper Symmetric Successive Overrelaxation (LU-SSOR) scheme~\cite{YOO87} using $4$ relaxation cycles, increasing the $\text{CFL}$ number after $100$ iterations to $\text{CFL}=50$;
\item A multigrid approach for the Navier-Stokes system over two grid levels with two iterations on the coarser grid;
\item A single fine level iteration for the turbulent equation alternating with a multigrid iteration for the RANS system.
\end{itemize}
Once the numerical parameters have been fixed, the number of iterations is determined from the evolution of the resulting global forces. A number of $2000$ iterations (the discrete residuals of all equations and their decrease being checked at every iteration) appeared to be acceptable. Hence this number of iterations has been retained for all calculations so far. Further discussions on this issue are available in Dumont \emph{et al.}~\cite{Onera14} (available on demand).

\subsection{Definition of the uncertainties}\label{sec:PDFs}

The aim of this work is to characterize the influence of uncertainties on the free-stream Mach number $M_\infty$, angle of attack $\alpha$, and thickness-to-chord ratio $r=h/c$ on some aerodynamic quantities of interest, such as the drag, lift, or pitching moment coefficients $C_D$, $C_L$, or $C_M$, respectively. These variable parameters are assumed to be independent and to follow Beta laws of the first kind $\beta_{\mathrm I}$. Therefore their probability density functions (PDF) read:
\begin{displaymath}
\beta_{\mathrm I}(x;a,b)={\mathbbm 1}_{[X_m,X_M]}(x)\frac{\Gamma(a+b)}{\Gamma(a)\Gamma(b)}\frac{(x-X_m)^{a-1}(X_M-x)^{b-1}}{(X_M-X_m)^{a+b-1}}\,.
\end{displaymath}
In the above $\Gamma(z)=\smash{\int_0^{+\infty}t^{z-1}\iexp^{-t}\id t}$ is the usual Gamma function, and $\smash{[X_m,X_M]}$ is the compact support of the random parameter $X\sim\smash{\beta_{\mathrm I}}$. The parameters $a=b$ as well as the bounds $\smash{X_m,X_M}$ for the three variable parameters $\xigj_1=r$, $\xigj_2=M_\infty$, $\xigj_3=\alpha$ are gathered in the Table \ref{tb:PDFs} below. This definition of uncertainties is part of the FP7 UMRIDA Project (\href{http://www.umrida.eu}{\texttt{www.umrida.eu}}), which gathers a novel data base of industrial challenges with prescribed uncertainties for the validation of UQ techniques against this series of relevant industrial test cases \cite{Airbus15}. We note in passing that the $\smash{\beta_{\mathrm I}}$ model is the one arising from Jaynes' maximum entropy principle \cite{JAY57a,JAY57b} when constraints on (i) the boundedness of the support $\smash{[X_m,X_M]}$, and (ii) the values of the averages $\smash{\esp{\log(X-X_m)}}$ and $\smash{\esp{\log(X_M-X)}}$, are imposed. Here and in the subsequent developments the usual notation $\esp{f(X)}=\smash{\int_\Rset f(x)P_X(\id x)}$ for $X\sim\smash{P_X}$ is employed.

\begin{table}[h!]
\begin{center}
\begin{tabular}{|c||c|c|c|c|}
\hline
& \makebox[3em]{$a=b$} &  \makebox[3em]{$X_m$} & \makebox[3em]{$X_M$} \\
\hline\hline
$\xigj_1$ & $4$ & $0.97\times\nominal{r}$ & $1.03\times\nominal{r}$ \\
$\xigj_2$ & $4$ & $0.95\times\nominal{M}_\infty$ & $1.05\times\nominal{M}_\infty$ \\
$\xigj_3$ & $4$ & $0.98\times\nominal{\alpha}$ & $1.02\times\nominal{\alpha}$ \\
\hline
\end{tabular}
\end{center}
\caption{Symmetric $\beta_{\mathrm I}$ laws for the variable geometrical and operational parameters.}\label{tb:PDFs}
\end{table}

\section{Polynomial surrogates}\label{sec:gPC}

The computation of the first moments (mean, standard deviation, skewness, kurtosis...) of the aerodynamic quantities of interest when the variability of the parameters above is accounted for, is done thank to surrogate models, or response surfaces. We more particularly focus on polynomial surrogates in this study.

Let $\model$ be a generic physical model involving $D$ parameters $\xig=(\xigj_1,\xigj_2,\dots\xigj_D)\in{\mathcal I}\subseteq\Rset^D$, such that the quantity of interest $\QoI\in\QoIset$ is given as:
\begin{displaymath}
\QoI=\model(\xig)\,.
\end{displaymath}
Let $\Pset^\torder[\xg]$ be the set of $D$-dimensional polynomials with total order $p$ with respect to $\xg\in\Rset^D$. We first note that this set has cardinality $\#\Pset^\torder[\xg]=\cardinal+1$ such that:
\begin{equation}\label{eq:cardinal}
\cardinal+1=\frac{(\torder+D)!}{\torder!D!}\,.
\end{equation}
A polynomial surrogate model $\surr{\model}_\torder$ of order $\torder$ for the model $\model$ is obtained as:
\begin{equation}\label{eq:optim0}
g\approx\surr{\model}_\torder=\arg\min_{\pi\in\Pset^\torder[\xg]}\demi\int_{\Rset^D}|g(\xg)-\pi(\xg)|^2\id P_\Xig(\xg)\,,
\end{equation}
where $\Xig\sim P_\Xig$ is the marginal PDF of the random parameters $\Xig$ with values in ${\mathcal I}\subseteq\smash{\Rset^D}$. The accuracy of this approximation may be assessed considering the limit of the mean-square norm $\smash{\esp{|\surr{\model}_\torder(\Xig)-\model(\Xig)|^2}}$ as $\torder\rightarrow +\infty$. However such a "convergence" does not necessarily holds, and it depends on the probability measure $P_\Xig$.

There are otherwise several ways to construct response surfaces and surrogates: embedded projection (this is the original spectral stochastic finite element method of Sun~\cite{SUN79} and Ghanem \& Spanos~\cite{Ghanem:2003} which is highly intrusive), non-intrusive projection (or collocation) \cite{GHI02,XIU05}, kriging \cite{krige:1951,kleijnen:2009,BOM11}, \emph{etc}.; see also Le Ma\^{\i}tre \& Knio~\cite{OLM:2010} for a detailed introduction. Regression is also an alternative, whereby an $\ell_2$ optimization problem is formed. For that purpose, a set of sampling points is needed in order to discretize the minimization problem (\ref{eq:optim0}). It is assumed in the following that these points are first chosen as the $\nquad$ integration points of a cubature rule $\smash{\cubature^\nquad}=\smash{\{\weight_\ell,\xig_\ell\}_{\ell=1}^\nquad}$, which provides with positive weights $\smash{\{\weight_\ell\}_{\ell=1}^\nquad}$ and nodes $\smash{\{\xig_\ell\}_{\ell=1}^\nquad}$ in $\smash{\Rset^D}$ such that for a smooth function $\xg\mapsto f(\xg)$ one can evaluate its average by:
\begin{equation}\label{eq:cubrule}
\int_{\mathcal I}f(\xg)\,\id P_\Xig(\xg)\simeq\sum_{\ell=1}^\nquad\weight_\ell f(\xig_\ell)\,.
\end{equation}

\subsection{Regression approach}

Using the foregoing cubature rule, the regression approach is formulated as a weighted least-squares minimization problem for the coefficients  $\{\coefj^\nquad_j\}_{j=0}^\cardinal$ of the polynomial surrogate $\surr{\model}_\torder^\nquad$ of $\model$ expanded on the monomials $\smash{\{[\xg]^j\}_{j=0}^\torder}$ of partial total order $j$ (\emph{i.e.} $\smash{[\xg]^j}=\smash{\xgj_1^{j_1}\xgj_2^{j_2}\cdots\xgj_D^{j_D}}$ with $j_1+j_2+\cdots+j_D\leq j$). Let $\coefv^\nquad=(\coefj_0^\nquad,\coefj_1^\nquad,\dots\coefj_\cardinal^\nquad)^\itr$ where $\cardinal$ is given by \eref{eq:cardinal}, then:
\begin{displaymath}
{\bf c}^\nquad=\arg\min_{{\boldsymbol d}\in\Rset^{\cardinal+1}}\demi\left(\vQoI-\mmeasure{\boldsymbol d}\right)^\itr\mweight\left(\vQoI-\mmeasure{\boldsymbol d}\right)\,,
\end{displaymath}
where $\vQoI=\smash{\{\QoI_\ell=g(\xig_\ell)\}_{\ell=1}^\nquad}$, $\smash{[\mmeasure]_{\ell j}=[\xig_\ell]^j}$, and $\mweight=\operatorname{diag}\{\weight_\ell\}_{\ell=1}^\nquad$; also ${\boldsymbol a}^\itr$ stands for the transpose of ${\boldsymbol a}$. Numerous methods are available to solve this problem whenever $\nquad\geq\cardinal+1$; we do not follow this approach in the subsequent developments though. We are rather interested in the situation whereby $\nquad\leq\cardinal+1$, and more interestingly $\nquad\ll\cardinal$. It is addressed subsequently in the section \ref{sec:CS}.

\subsection{Projection approach}

We assume now that a polynomial basis $\basis$ of $L^2({\mathcal I},P_\Xig)$ is available. Then we construct the polynomial surrogate $\surr{\model}_\torder$ of $\model$ by standard $L^2$ projection on the finite dimensional subspace of $L^2({\mathcal I},P_\Xig)$ spanned by the truncated family of orthonormal polynomials up to the total order $\torder$ denoted by $\smash{\basis^\torder}=\smash{\{\vbase_j\}_{j=0}^\cardinal}$, where $\cardinal$ is again given by \eref{eq:cardinal}. The orthonormalization of this basis reads:
\begin{equation}\label{eq:orthonorm}
\int_{\mathcal I}\vbase_j(\xg)\vbase_k(\xg)\id P_\Xig(\xg)=(\vbase_j,\vbase_k)_{L^2}=\delta_{jk}\,.
\end{equation}
Then $\surr{\model}_\torder=\sum_{j=0}^\cardinal\model_j\vbase_j$ where $\model_j=(\model,\vbase_j)_{L^2}$, $0\leq j\leq\cardinal$. Using the cubature rule of \eref{eq:cubrule}, these expansion coefficients are approximated by:
\begin{displaymath}
\model_j\approx\model_j^\nquad=\sum_{\ell=1}^\nquad\weight_\ell\QoI_\ell\vbase_j(\xig_\ell)\,,\quad 0\leq j\leq\cardinal\,.
\end{displaymath}
This corresponds to the approximation $\smash{\surr{\model}_\torder\approx\surr{\model}_\torder^\nquad=\sum_{j=0}^\cardinal\model_j^\nquad\vbase_j}$. Such representations are referred to as polynomial chaos (PC) expansions in the dedicated literature, provided that the variable parameters $\Xig$ follow a multi-dimensional Gaussian (normal) distribution $\smash{P_\Xig}=\smash{\bigotimes_{d=1}^D{\mathcal N}(0,1)}$ \cite{Ghanem:2003,OLM:2010}. They are otherwise called generalized polynomial chaos (gPC) expansions for other distributions \cite{XIU02,SOI04}.

\subsection{Application to Uncertainty Quantification (UQ)}

Once the polynomial surrogate model $\surr{\model}_\torder^\nquad$ has been derived, the first moments and/or cumulants of the quantity of interest $\QoI$ can be computed using the cubature rule $\cubature^\nquad$ and evaluations $\{\QoI_\ell\}_{\ell=1}^\nquad$ of the physical model $\model$ at these nodes. Indeed, for a regular function $\QoI\mapsto f(\QoI)$ on $\QoIset$ one can estimate a mean output functional by:
\begin{displaymath}
\esp{f(\QoI)}=\int_{\mathcal I}f(g(\xg))\id P_\Xig(\xg)\simeq\sum_{\ell=1}^\nquad\weight_\ell f(\QoI_\ell)\,.
\end{displaymath}
The mean $\mu$ is obtained for $f(\QoI)=\QoI$, the variance $\sigma^2$ is obtained for $f(\QoI)=\smash{(\QoI-\mu)^2}$, the skewness $\gamma_1$ for $f(\QoI)=\smash{(\frac{\QoI-\mu}{\sigma})^3}$, the kurtosis $\beta_2$ for $f(\QoI)=\smash{(\frac{\QoI-\mu}{\sigma})^4}$, \emph{etc}. More generally, the $j$-th moment $\smash{m_j}$ is obtained for $f(\QoI)=\smash{\QoI^j}$, and may be used to compute the characteristic function $\smash{\Phi_\QOI}$:
\begin{displaymath}
\Phi_\QOI(u)=\int_\QoIset\iexp^{\ci u\cdot\QoI}\id P_\QOI(\QoI)=\sum_{j=0}^{+\infty}\frac{m_j}{j!}(\ci u)^j\,,
\end{displaymath}
where by the causality principle (or transport of PDFs) for $\QOI\sim g(\Xig)$ one has:
\begin{displaymath}
P_\QOI(\id\QoI)=\left|\frac{\id\model^{-1}}{\id\QoI}\right|P_\Xig(\model^{-1}(\id\QoI))\,.
\end{displaymath}

Sensitivity indices may be computed alike. Denoting by ${\mathscr I}_d$ the set of indices corresponding to the polynomials of $\basis^p$ depending only on the $d$--th variable parameter $\xigj_d$, the main-effect gPC-based Sobol' indices are given by (see \emph{e.g.} Sudret~\cite{SUD08}):
\begin{displaymath}\label{eq:main_sensitivity}
S_d=\frac{1}{\sigma^2}\sum_{j\in{\mathscr I}_d}g_j^2\,,
\end{displaymath}
owing to the normalization condition (\ref{eq:orthonorm}). More generally, if $\smash{{\mathscr I}_{d_1d_2\dots d_s}}$ is the set of indices corresponding to the polynomials of $\basis^p$ depending only on the parameters $\smash{\xigj_{d_1},\xigj_{d_2},\dots\xigj_{d_s}}$, the $s$--fold joint sensitivity indices are:
\begin{displaymath}
S_{d_1d_2\dots d_s}=\frac{1}{\sigma^2}\sum_{j\in{\mathscr I}_{d_1d_2\dots d_s}}g_j^2\,.
\end{displaymath}
In the subsequent application to the two-dimensional configuration described in section~\ref{sec:BC02}, we will consider the main-effect sensitivity indices $S_d$ and the $2$--fold joint sensitivity indices $\smash{S_{d_1d_2}}$.

\section{Application to the two-dimensional RAE 2822 transonic airfoil}\label{sec:gPC-BC02}

From the previous analysis, we see that the main ingredients requested for the construction of polynomial surrogates are the cubature rule $\cubature^\nquad$ and the truncated polynomial basis $\basis^\torder$, for $\nquad$ integration nodes and a multi-dimensional polynomial total order $\torder$. In addition we have here $D=3$ for the parameter space dimension. Owing to the choices made for the variable parameters considered for this case (see Table \ref{tb:PDFs}), we have $\xig=\smash{(\xigj_1,\xigj_2,\xigj_3)}\in\smash{\prod_{d=1}^3[X_m^{(d)},X_M^{(d)}]}$ and $\smash{P_\Xig}=\smash{\bigotimes_{d=1}^3\beta_{\mathrm I}(4,4)}$. Therefore the integration points should be chosen from a Gauss-Jacobi quadrature rule, and the polynomial basis should be constituted by the multi-dimensional Jacobi polynomials which are orthogonal with respect to the weight function $\xg\mapsto\smash{w(\xg)=\prod_{d=1}^3(1-\xgj_d^2)^3}$.

\subsection{Polynomial basis}

The polynomial basis $\smash{\basis^\torder}$ adapted to the parameters PDF $P_\Xig$ is constituted by the three-dimensional Jacobi polynomials $\smash{\vbase_\vj}$, $\smash{\vj=(j_1,j_2,j_3)\in\Nset^3}$, such that $\smash{\|\vj\|_1=j_1+j_2+j_3\leq\torder}$. They are computed by:
\begin{displaymath}
\vbase_\vj(\xg)=\prod_{d=1}^3\vbase_{j_d}(\xgj_d)\,,\quad\|\vj\|_1\leq\torder\,,
\end{displaymath}
where $\smash{\{\vbase_{j_d}\}_{j_d\geq 0}}$ is the family of one-dimensional orthonormal Jacobi polynomials with respect to the weight function $\xgj\mapsto w_1(\xgj)=(1-\xgj^2)^3$.

In the present study the polynomial surrogates $\surr{\model}_\torder$ constructed for the evaluation of the drag, lift and pitching moment coefficients $C_D$, $C_L$ and $C_M$, respectively, are truncated up to the total order $\torder=8$. Therefore $\smash{\cardinal+1={\tiny\begin{pmatrix} \torder+3\\3\end{pmatrix}}}=165$ multi-dimensional Jacobi polynomials are ultimately retained in those gPC expansions.

\subsection{Cubature rules}

\begin{figure}
\centering{\includegraphics[width=7.4cm]{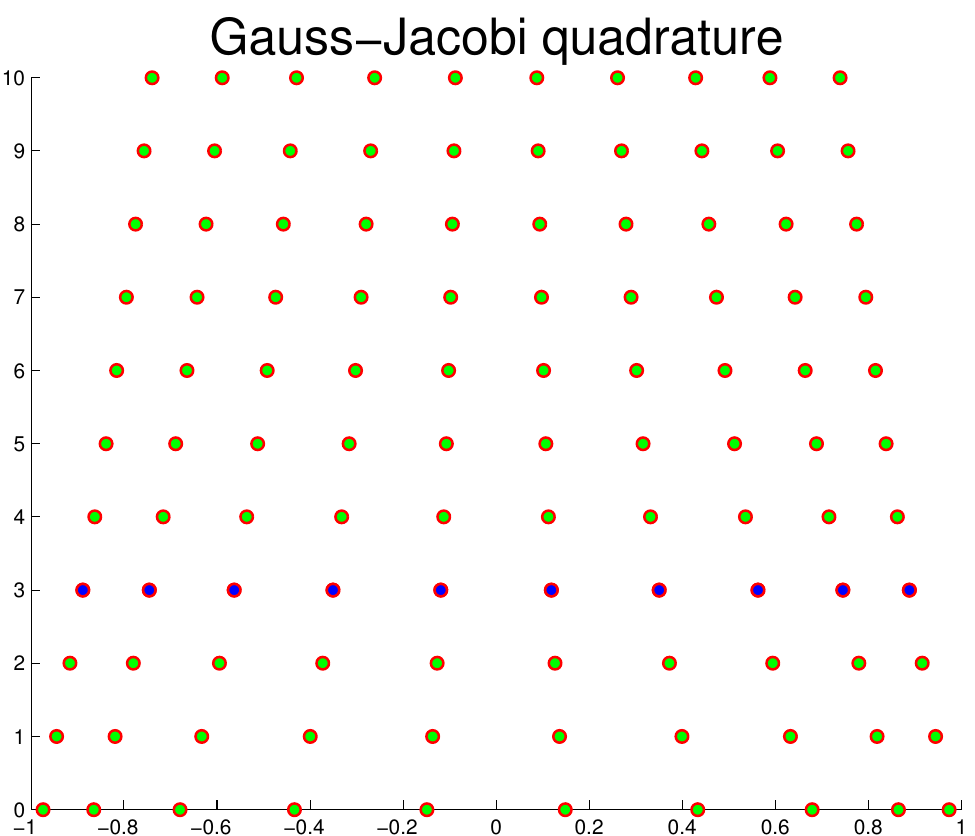}}
\caption{Gauss-Jacobi quadratures for $0\leq a=b\leq 10$, $N=10$ and $N_b=0$. The blue dots correspond to $a=b=3$ for which the Jacobi weight function $\xgj\mapsto w^{(3,3)}(\xgj)$ is identified with the $\beta_{\mathrm I}(4,4)$ PDF up to a normalization constant.}\label{fg:GJrules}
\end{figure}

\begin{figure}
\centering{\includegraphics[width=7.4cm]{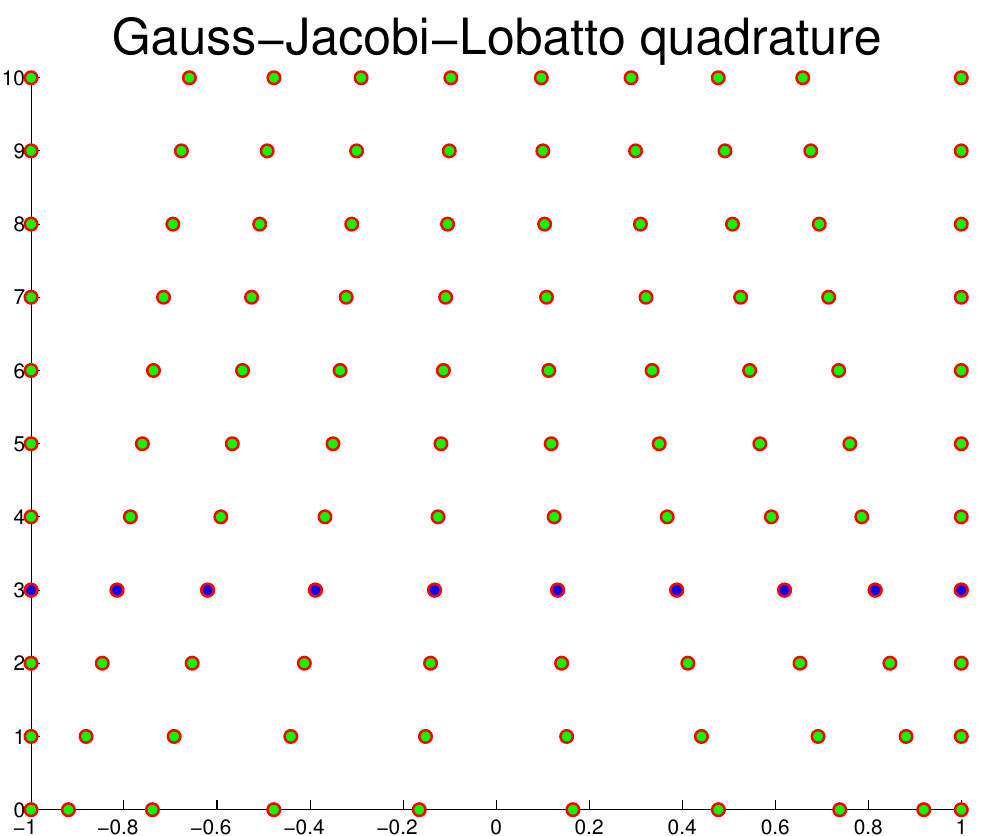}}
\caption{Gauss-Jacobi quadratures for $0\leq a=b\leq 10$, $N=10$ and $N_b=2$. The blue dots correspond to $a=b=3$ for which the Jacobi weight function $\xgj\mapsto w^{(3,3)}(\xgj)$ is identified with the $\beta_{\mathrm I}(4,4)$ PDF up to a normalization constant.}\label{fg:GJLrules}
\end{figure}

One-dimensional Gauss-Jacobi quadratures $\Theta_1^\level$ based on $\level$ integration points are tailored to integrate on $[-1,1]$ a smooth function $\xgj\mapsto f(\xgj)$:
\begin{equation}\label{eq:GJrule}
\int_{-1}^1f(\xgj)(1-\xgj)^a(1+\xgj)^b\id\xgj\simeq\sum_{\ell=1}^{\level-N_b}\weight_\ell f(\xigj_\ell)+\sum_{\ell'=1}^{N_b}\weight_{N-N_b+\ell'} f(\xigj_{N-N_b+\ell'})\,,\quad a,b>-1\,,
\end{equation}
such that this rule turns to be exact for polynomials up to the order $2N-1-N_b$. Here $N_b$ is the number of fixed nodes of the rule, typically the bounds $\pm 1$. Depending on the choice of $N_b$, different terminologies are used: 
\begin{itemize}
\item $N_b=0$ is the classical Gauss-Jacobi rule;
\item $N_b=1$ is the Gauss-Jacobi-Radau (GJR) rule, choosing $\xigj_N=-1$ or $\xigj_N=1$ for instance;
\item $N_b=2$ is the Gauss-Jacobi-Lobatto (GJL) rule, choosing $\xigj_{N-1}=-1$ and $\xigj_N=1$ for instance.
\end{itemize}
Since in our case we have chosen a total order $\torder=8$, $\level=10$ GJL points are needed to recover exactly the orthonormality property (\ref{eq:orthonorm}) for the corresponding one-dimensional Jacobi polynomials. Indeed $N$ should be defined such that $2N-3\geq 16$ in this situation. The $10$--points Gauss-Jacobi rules are illustrated on \fref{fg:GJrules} for various values of the parameters $a=b$ of the Jacobi weight function $\xgj\mapsto w^{(a,b)}(\xgj)=(1-\xgj)^a(1+\xgj)^b$, and the $10$--points Gauss-Jacobi-Lobatto rules are illustrated on \fref{fg:GJLrules}. The blue dots correspond to $a=b=3$ and thus pertain to the $\beta_{\mathrm I}(4,4)$ PDF. The reason why we include the boundary nodes in the quadrature rule stems from the fact that the basic engineering practice would consider the evaluation of the physical model $\model$ at the bounds of the support of the variable parameters. The main advantage of using Gauss-Jacobi quadratures is that they do not add integration points for the increased order $a+b$ induced by the weight function $w^{(a,b)}$. The Clenshaw-Curtis rule \cite{CleCur_60} for example is typically suited for polynomials of order $N-1$, yet higher orders are actually achieved in practice \cite{Tre_08}. Thus if one uses $N$ nodes from this rule to compute the left-hand side of (\ref{eq:GJrule}) an exact evaluation is achieved for polynomials up to the order $N-1-(a+b)$, instead of $2N-3$ with a GJL rule. However, the latter does not have the nesting property of the former.

\begin{figure}
\centering{\includegraphics[width=7.4cm]{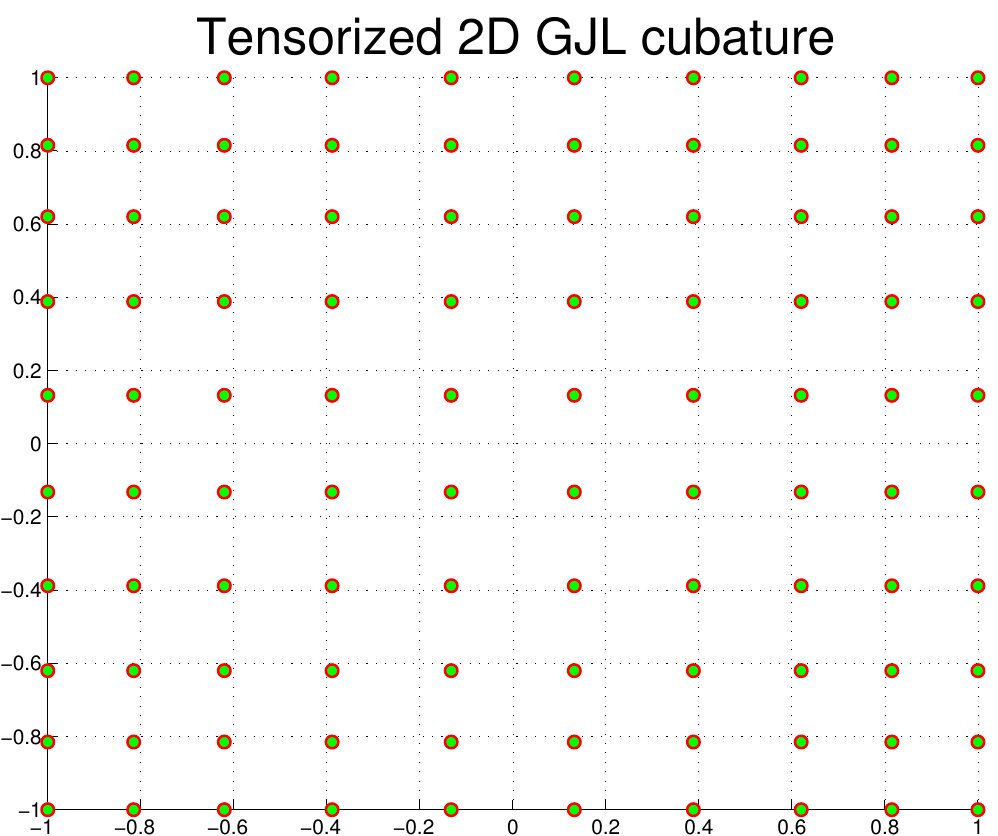}}
\caption{$10$--th level tensorized two-dimensional GJL cubatures for $a=b=3$ ($\nquad=10^2$).}\label{fg:2D-10levelT}
\end{figure}

\begin{figure}
\centering{\includegraphics[width=7.4cm]{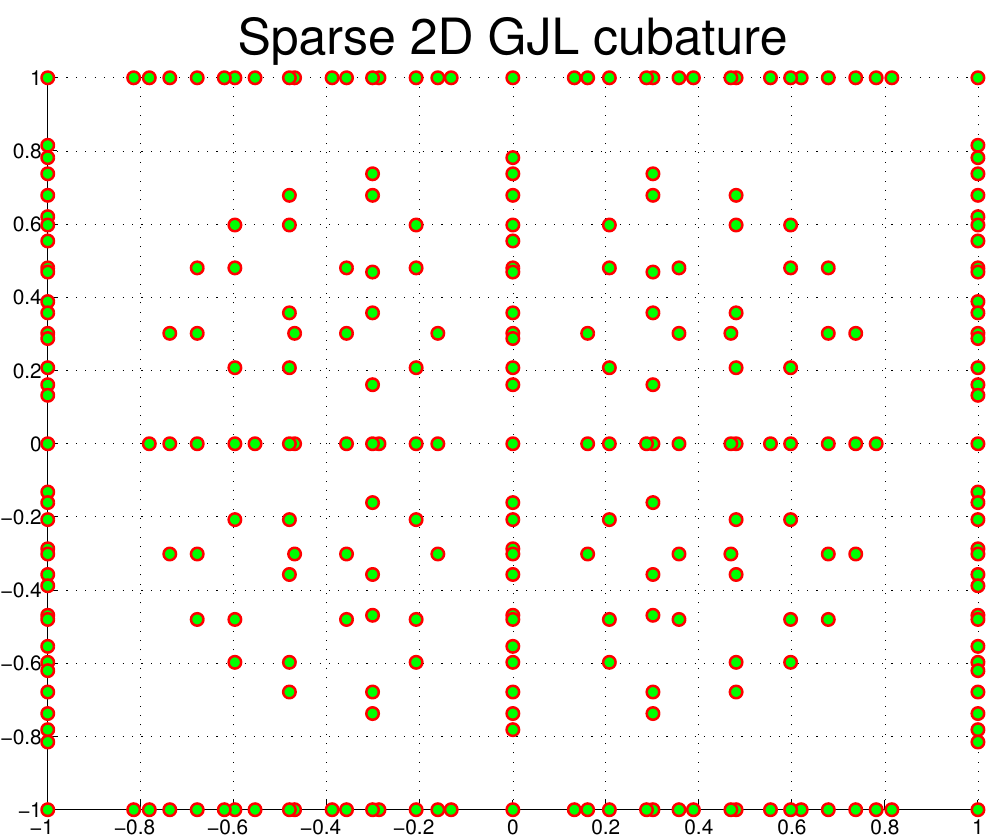}}
\caption{$10$--th level sparse two-dimensional GJL cubatures for $a=b=3$ ($\nquad=273$).}\label{fg:2D-10levelS}
\end{figure}

\begin{figure}
\centering{\includegraphics[width=7.4cm]{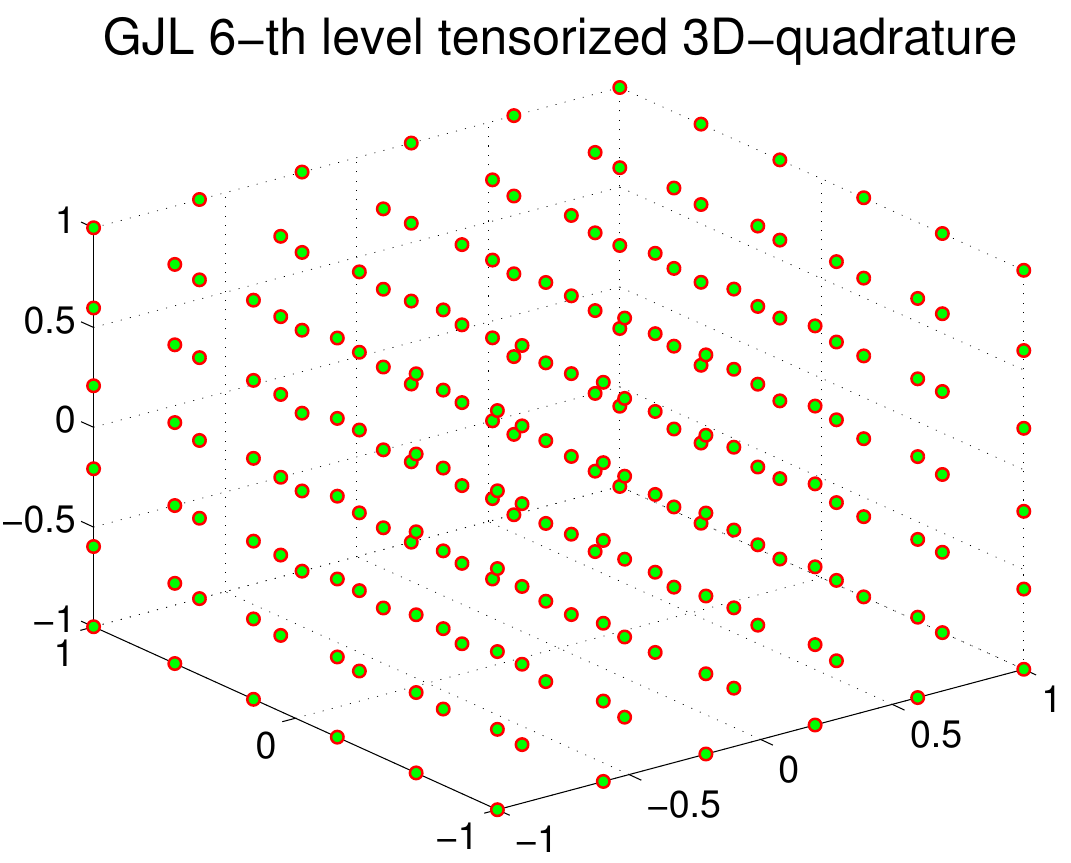}}
\caption{$6$--th level tensorized three-dimensional GJL cubatures for $a=b=3$ ($\nquad=6^3=216$).}\label{fg:3D-6levelT}
\end{figure}

\begin{figure}
\centering{\includegraphics[width=7.4cm]{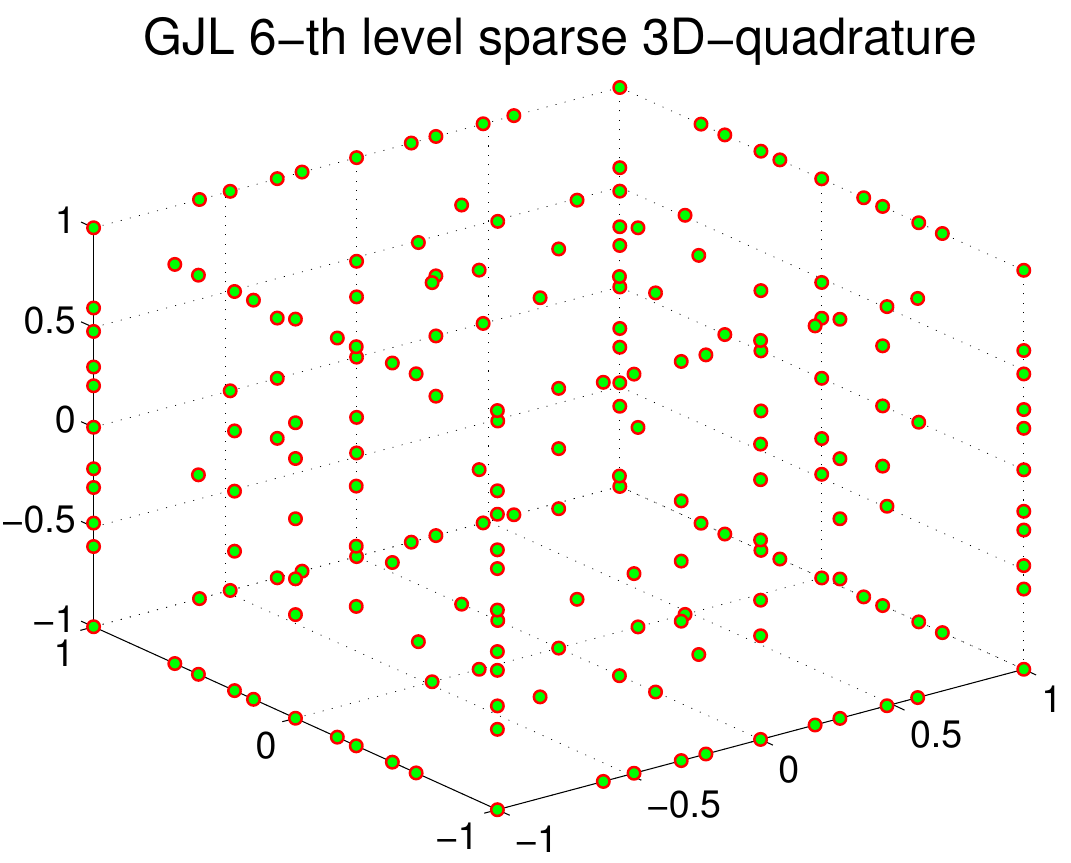}}
\caption{$6$--th level sparse three-dimensional GJL cubatures for $a=b=3$ ($\nquad=201$).}\label{fg:3D-6levelS}
\end{figure}

Multi-dimensional quadratures (cubatures) may subsequently be obtained by tensorization and/or sparsification of these one-dimensional rules. Firstly, a multi-dimensional tensorized grid is obtained by the straightforward product rule:
\begin{equation}\label{eq:tensorcub}
\cubature^\nquad=\bigotimes_{d=1}^D\Theta_1^\level\,.
\end{equation}
It contains $\nquad=\level^D$ grid points, \emph{i.e.} $\nquad=1,000$ for the present Test Case ($\level=10$, $D=3$). Of course the order $N$ could be adapted depending on the $d$--th variable parameter. Secondly, a sparse cubature rule can be derived thank to the Smolyak algorithm \cite{SMO63}. The so-called $\level$--th level, $D$-dimensional sparse grid $\cubature_{D,\level}$ is obtained by the following linear combination of product formulas for the nodes~\cite{WAS95}:
\begin{equation}\label{eq:sparsecub}
\cubature_{D,\level}=\bigcup_{\level+1\leq\|\vj\|_1\leq \level+D}\Theta_1^{j_1}\otimes\cdots\otimes\Theta_1^{j_D}\,.
\end{equation}
For example, if $\level=5$ and $D=3$ one has $\nquad=99$ for the total number of grid points using a one-dimensional GJL quadrature $\smash{\Theta_1^j}$ as the generating rule, and:
\begin{displaymath}
\cubature_{3,5}=\Theta_1^2\otimes\Theta_1^2\otimes\Theta_1^2+\Theta_1^2\otimes\Theta_1^2\otimes\Theta_1^3+\Theta_1^2\otimes\Theta_1^3\otimes\Theta_1^3 + \text{perm.}
\end{displaymath}
By a direct extension of the arguments devised by Novak \& Ritter~\cite{NOV99} or Heiss \& Winschel~\cite{HEI08}, it can be shown that such a $\level$--th level, $D$-dimensional sparse rule based on the GJL one-dimensional rule is exact for $D$-dimensional polynomials of total order $2\level-3$. The total number of integration points of the rule is given for $D\gg 1$ and $\level$ fixed by the estimate $\nquad=\smash{\mathrm{O}(\smash{\frac{(2D)^\level}{\level!}})}$ (see \emph{e.g.} Novak \& Ritter~\cite{NOV99} and references therein; the dual estimate for $D$ fixed and $\level\gg 1$ is $\nquad=\smash{\mathrm{O}(\smash{\frac{(2\level)^D}{D!}})}$), which typically outperforms the product rule with $n=\level^D$ for $D\geq 4$. It is gathered in the Table \ref{tb:sparseGJLrule} below for various combinations $(D,\level)$. The sparse rule is plotted in \fref{fg:2D-10levelS} for $\level=10$ and $D=2$, and in \fref{fg:3D-6levelS} for $\level=6$ and $D=3$ together with the corresponding product rule in \fref{fg:2D-10levelT} and \fref{fg:3D-6levelT}, respectively, for illustration purposes. We note that since the underlying one-dimensional GJL rule is not nested, the multi-dimensional rules are not either. Also the weights of the latter may be negative for some nodes although the underlying one-dimensional rules have positive weights.

\begin{table}[h!]
\begin{center}
\begin{tabular}{|c||c|c|c|c|c|}
\hline
\backslashbox{$\level$}{$D$}  & \makebox[3em]{2} &  \makebox[3em]{3} & \makebox[3em]{4} & \makebox[3em]{5} & \makebox[3em]{6} \\
\hline\hline
2 & 4 & 8 & 16 & 32 & 64 \\
3 & 8 & 20 & 48 & 112 & 256 \\
4 & 17 & 50 & 136 & 352 & 880 \\
5 & 29 & 99 & 304 & 872 & 2384 \\
6 & 53 & 201 & 673 & 2082 & 6092\\
7 & 85 & 363 & 1337 & 4483 & 14072\\
8 & 133 & 647 & 2585 & 9293 & 31025 \\
9 & 193 & 1079 & 4697 & 18143 & 64469 \\ 
10 & 273 & 1769 & 8321 & 34323 & 129197 \\
\hline
\end{tabular}
\end{center}
\caption{The total number of grid points for the $\level$--th level, $D$-dimensional sparse rule based on the GJL one-dimensional rule.}\label{tb:sparseGJLrule}
\end{table}

\subsection{Results}\label{sec:results_quad}

The first four moments of the drag, lift, and pitching moment coefficients $C_D$, $C_L$, and $C_M$, respectively, are gathered in the Table \ref{tb:tensorcub} below for $\nquad=10^3=1,000$ calls to the model $g$ using the \emph{elsA} software~\cite{CAM13,elsa} with the $10$--th level full product rule (\ref{eq:tensorcub}). Table \ref{tb:sparsecub} gathers the same results using a $6$-th level sparse rule (\ref{eq:sparsecub}) for which $\nquad=201$ from Table \ref{tb:sparseGJLrule}. The reasons why we have not used a $10$--th level sparse rule for a consistent comparison with the $10$--th level product rule are twofold. Firstly, for this case $\nquad=1769$, which is not competitive with the product rule. Secondly, the gPC coefficients are observed to be sparse so that the higher-order three-dimensional polynomials contribute only marginally to the surrogates. This observation is elaborated further on in the subsequent section \ref{sec:CS} dealing with the sparse reconstruction approach we have applied invoking the theory of compressed sensing. Using the $6$-th level sparse rule we are able to integrate exactly three-dimensional polynomials up to total order $9$, hence reconstruct polynomial surrogates $\surr{\model}_{\torder'}$ up to total order $\torder'=4$, corresponding to $\smash{\cardinal'+1={\tiny\begin{pmatrix} \torder'+3\\3\end{pmatrix}}}=35$ multi-dimensional Jacobi polynomials in those gPC expansions.

\begin{table}[h!]
\begin{center}
\begin{tabular}{|c||c|c|c|c|}
\hline
& \makebox[3em]{$\mu$} &  \makebox[3em]{$\sigma$} & \makebox[3em]{$\gamma_1$} & \makebox[3em]{$\beta_2$} \\
\hline\hline
$C_D$ & 133.37e-04 & 34.128e-04  & 1.0237e+00  & 3.3030e+00 \\
$C_L$ & 72.274e-02 & 1.6695e-02 & -9.6221e-01 &  3.0630e+00 \\
$C_M$ & -453.99e-04 &  32.239e-04 & -5.7845e-01 & 2.3190e+00 \\
\hline
\end{tabular}
\end{center}
\caption{First four moments of the aerodynamic coefficients computed by the $10$--th level product rule with $\nquad=10^3$.}\label{tb:tensorcub}
\end{table}

\begin{table}[h!]
\begin{center}
\begin{tabular}{|c||c|c|c|c|}
\hline
 & \makebox[3em]{$\mu$} &  \makebox[3em]{$\sigma$} & \makebox[3em]{$\gamma_1$} & \makebox[3em]{$\beta_2$} \\
\hline\hline
$C_D$ & 133.38e-04 & 34.097e-04  & 1.0293e+00 & 3.2611e+00 \\
$C_L$ &  72.269e-02 & 1.6729e-02  & -9.8175e-01 & 2.8678e+00 \\
$C_M$ & -453.96e-04 & 32.175e-04 & -5.9533e-01 & 2.3885e+00 \\
\hline
\end{tabular}
\end{center}
\caption{First four moments of the aerodynamic coefficients computed by the $6$--th level sparse rule with $\nquad=201$.}\label{tb:sparsecub}
\end{table}

The main-effect sensitivity indices computed with the $10$--th level product rule are gathered in Table \ref{tb:Soboli-tensor} below, while the joint sensitivity indices are gathered in Table \ref{tb:Sobolij-tensor}. Tables \ref{tb:Soboli-sparse} and \ref{tb:Sobolij-sparse} display the same indices computed with the $6$--th level sparse rule. It is clearly apparent from these results that the free-stream Mach number is the chief parameter controlling the variability of the aerodynamic coefficients in the range of analysis considered for this test case. We may also emphasize the discrepancies between the sensitivity indices computed with a full and a sparse rule, although their theoretical accuracy are different in the present case. These differences are even more pronounced for the joint sensitivity indices. The sparse reconstruction approach outlined in the next section yields sensitivities very comparable to the ones derived with the product rule, at a much lower computational cost though.

\begin{table}[h!]
\begin{center}
\begin{tabular}{|c||c|c|c|c|}
\hline
& \makebox[3em]{$\xigj_1=h/c$} &  \makebox[3em]{$\xigj_2=M_\infty$} & \makebox[3em]{$\xigj_3=\alpha$} \\
\hline\hline
$C_D$ & 0.081e-01 & 9.892e-01 &  0.008e-01  \\
$C_L$ & 0.034e-01 & 9.554e-01 & 0.286e-01 \\
$C_M$ & 0.269e-01 & 9.721e-01 & 0.000e-01 \\
\hline
\end{tabular}
\end{center}
\caption{Main-effect sensitivity indices of the aerodynamic coefficients computed by the $10$--th level product rule with $\nquad=10^3$.}\label{tb:Soboli-tensor}
\end{table}

\begin{table}[h!]
\begin{center}
\begin{tabular}{|c||c|c|c|c|}
\hline
& \makebox[3em]{$\xigj_2\xigj_3$} &  \makebox[3em]{$\xigj_1\xigj_3$} & \makebox[3em]{$\xigj_1\xigj_2$} \\
\hline\hline
$C_D$ &  0.021e-02  & 0.000e-02 &  0.172e-02 \\
$C_L$ &  0.036e-02 &  0.000e-02 &  1.221e-02 \\
$C_M$ & 0.007e-02  & 0.000e-02  &  0.089e-02 \\
\hline
\end{tabular}
\end{center}
\caption{Joint sensitivity indices of the aerodynamic coefficients computed by the $10$--th level product rule with $\nquad=10^3$.}\label{tb:Sobolij-tensor}
\end{table}

\begin{table}[h!]
\begin{center}
\begin{tabular}{|c||c|c|c|c|}
\hline
& \makebox[3em]{$\xigj_1=h/c$} &  \makebox[3em]{$\xigj_2=M_\infty$} & \makebox[3em]{$\xigj_3=\alpha$} \\
\hline\hline
$C_D$ &  0.052e-01 &  6.635e-01 & 0.007e-01 \\
$C_L$ &  0.033e-01 &  5.883e-01 & 0.195e-01 \\
$C_M$ & 0.256e-01 &  9.227e-01 & 0.002e-01 \\
\hline
\end{tabular}
\end{center}
\caption{Main-effect sensitivity indices of the aerodynamic coefficients computed by the $6$--th level sparse rule with $\nquad=201$.}\label{tb:Soboli-sparse}
\end{table}

\begin{table}[h!]
\begin{center}
\begin{tabular}{|c||c|c|c|c|}
\hline
& \makebox[3em]{$\xigj_2\xigj_3$} &  \makebox[3em]{$\xigj_1\xigj_3$} & \makebox[3em]{$\xigj_1\xigj_2$} \\
\hline\hline
$C_D$ &  0.138e-02 &  23.944e-02 & 0.847e-02 \\
$C_L$ &  0.287e-02 &  29.990e-02  & 1.510e-02 \\
$C_M$ & 0.085e-02 & 3.276e-02 & 0.582e-02 \\
\hline
\end{tabular}
\end{center}
\caption{Joint sensitivity indices of the aerodynamic coefficients computed by the $6$--th level sparse rule with $\nquad=201$.}\label{tb:Sobolij-sparse}
\end{table}

\begin{figure}
\centering{\includegraphics[width=7.4cm]{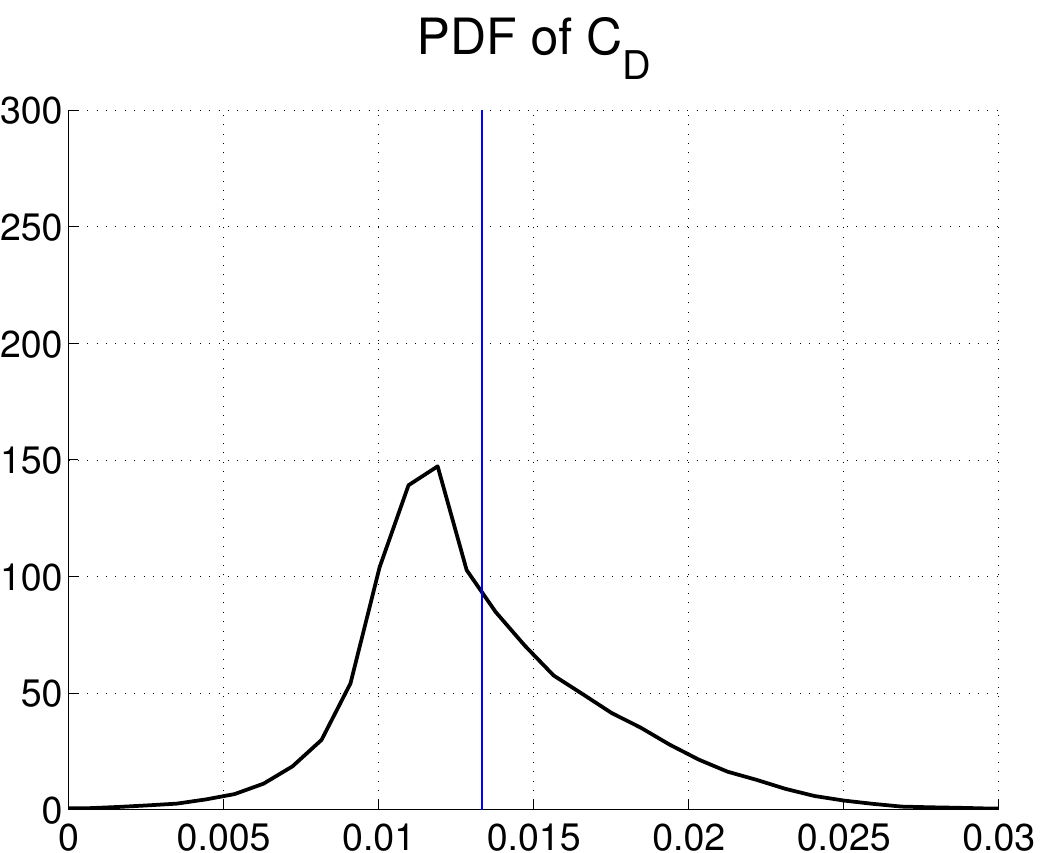}}
\caption{PDF of the drag coefficient $C_D$ computed by the $10$--th level product rule with $\nquad=10^3$.}\label{fg:PDFCD-tensor}
\end{figure}

\begin{figure}
\centering{\includegraphics[width=7.4cm]{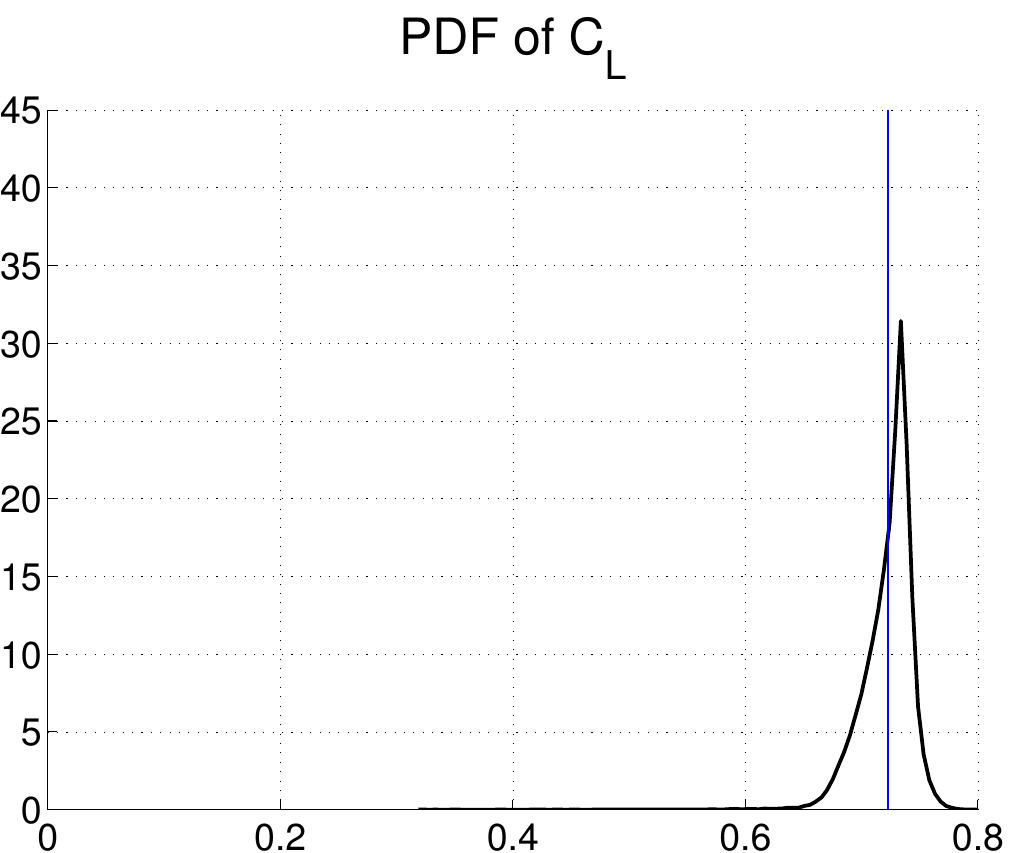}}
\caption{PDF of the lift coefficient $C_L$ computed by the $10$--th level product rule with $\nquad=10^3$.}\label{fg:PDFCL-tensor}
\end{figure}

\begin{figure}
\centering{\includegraphics[width=7.4cm]{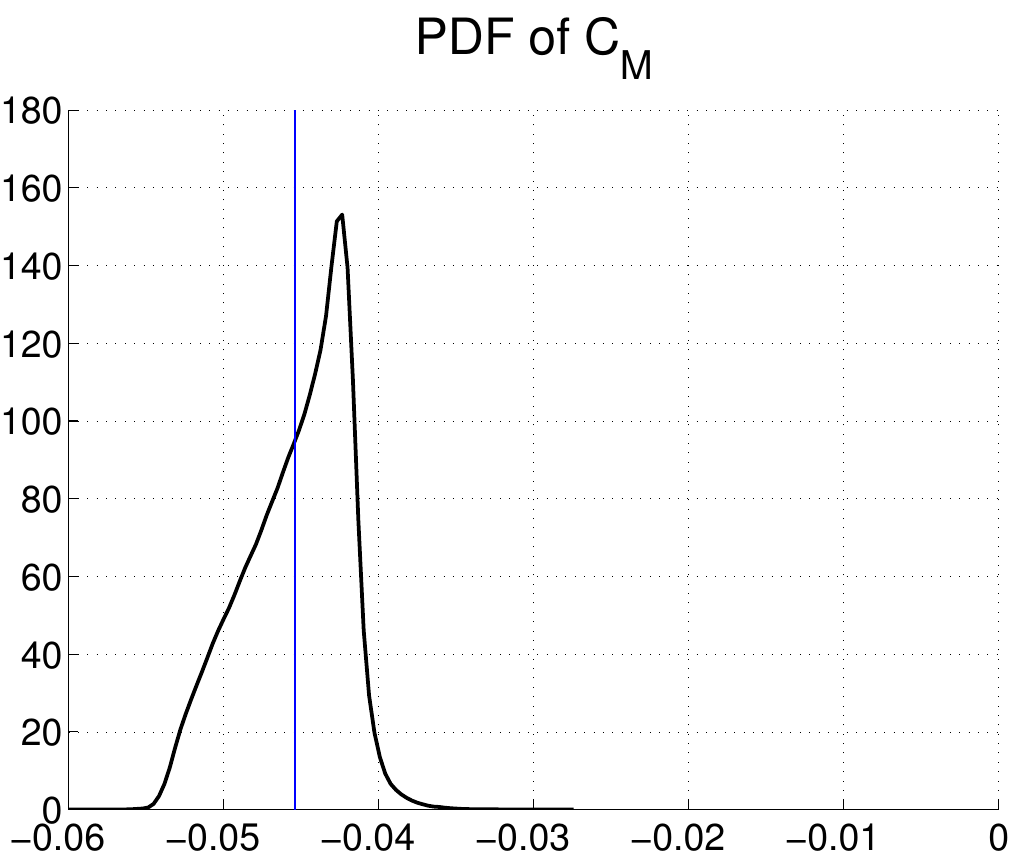}}
\caption{PDF  of the pitching moment coefficient $C_M$ computed by the $10$--th level product rule with $\nquad=10^3$.}\label{fg:PDFCM-tensor}
\end{figure}

\begin{figure}
\centering{\includegraphics[width=7.4cm]{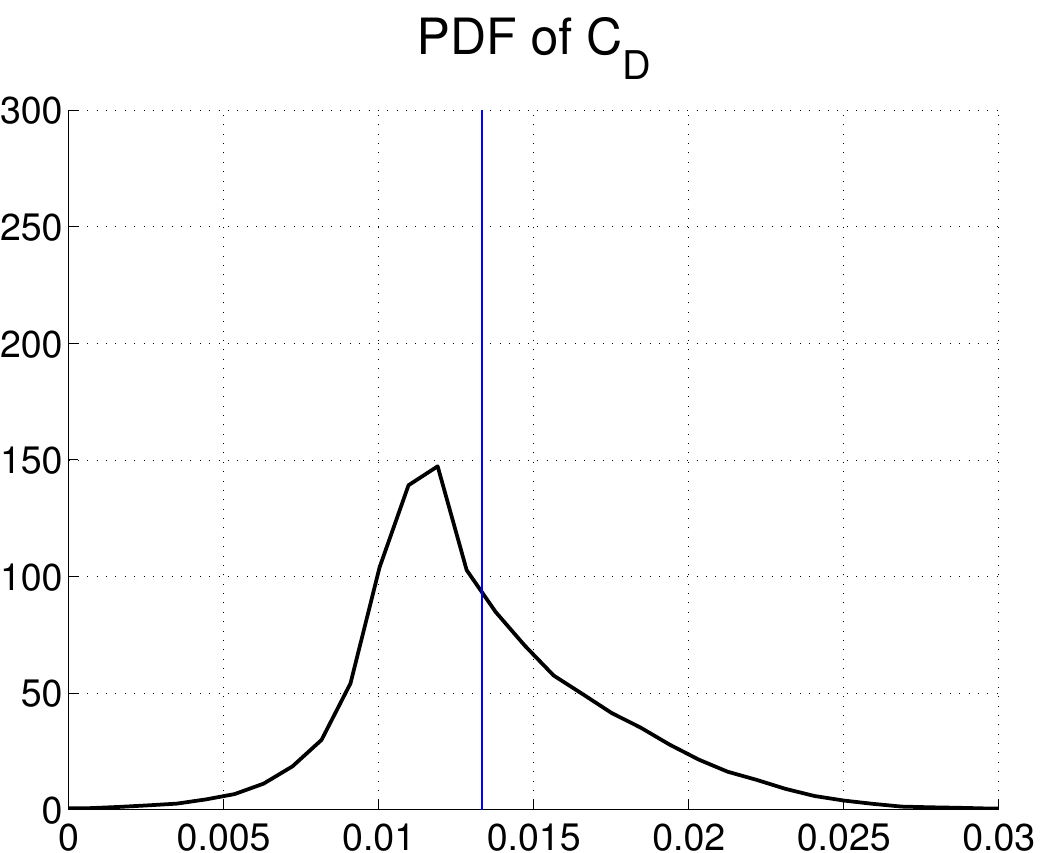}}
\caption{PDF of the drag coefficient $C_D$ computed by the $6$--th level sparse rule with $\nquad=201$.}\label{fg:PDFCD-sparse}
\end{figure}

\begin{figure}
\centering{\includegraphics[width=7.4cm]{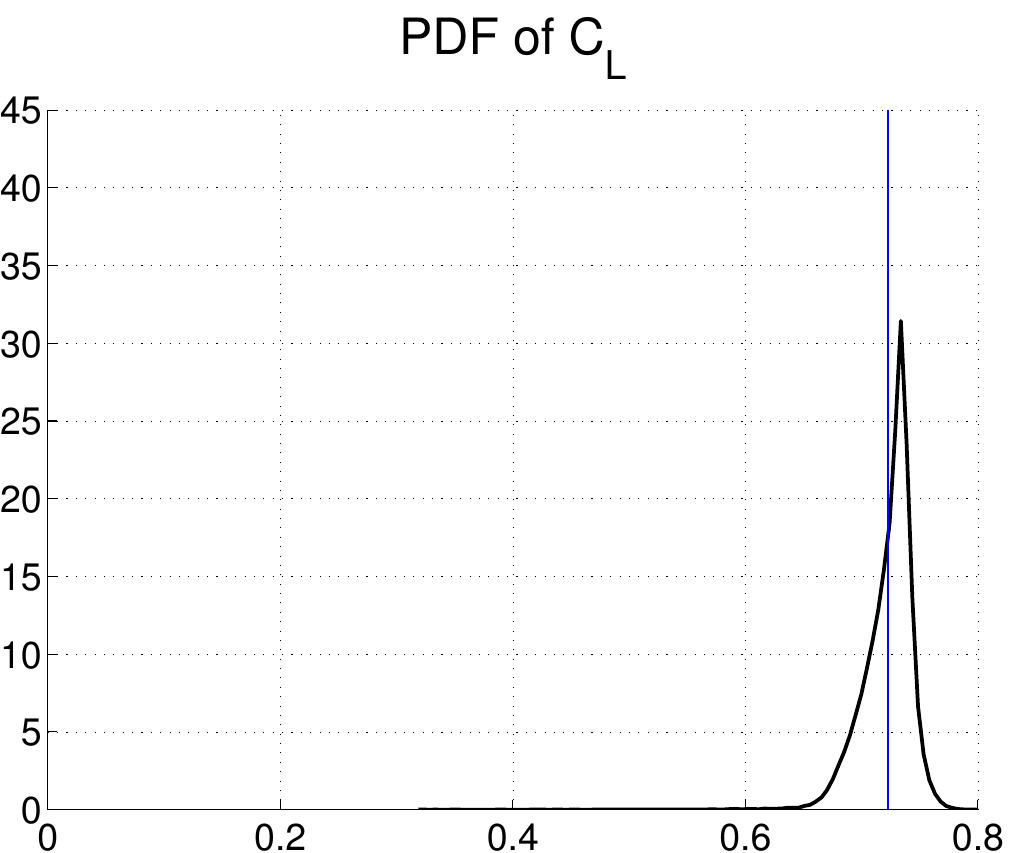}}
\caption{PDF of the lift coefficient $C_L$ computed by the $6$--th level sparse rule with $\nquad=201$.}\label{fg:PDFCL-sparse}
\end{figure}

\begin{figure}
\centering{\includegraphics[width=7.4cm]{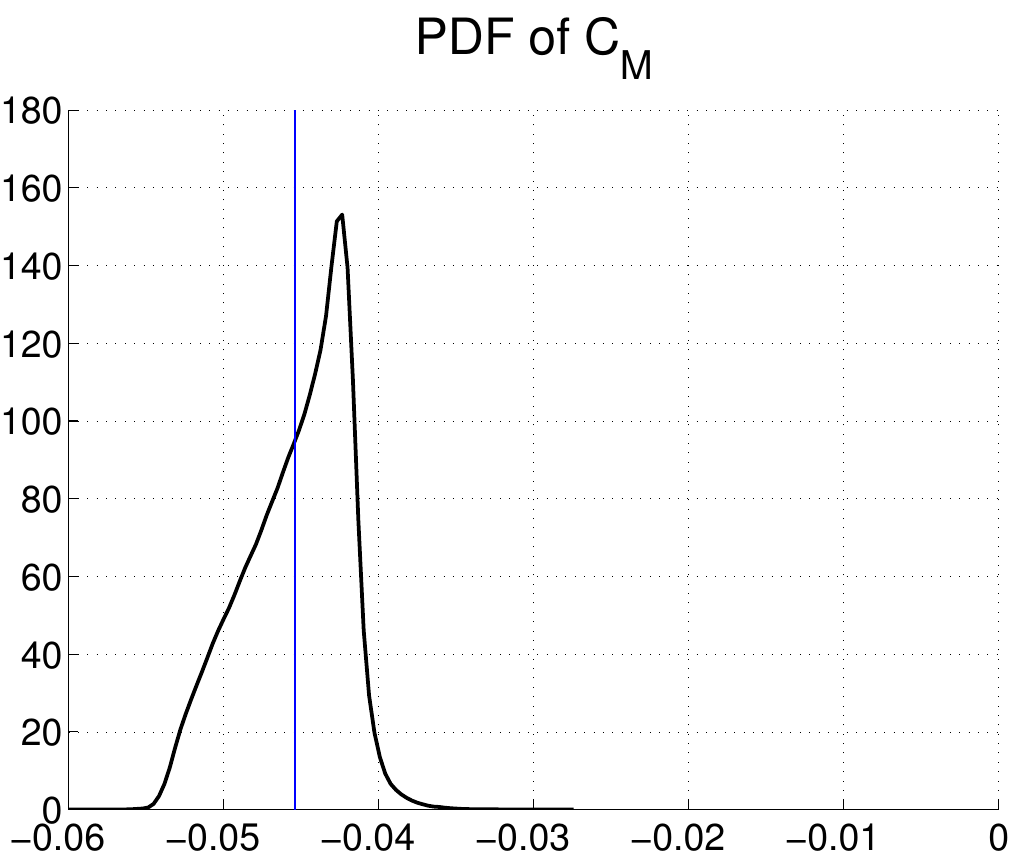}}
\caption{PDF of the pitching moment coefficient $C_M$ computed by the $6$--th level sparse rule with $\nquad=201$.}\label{fg:PDFCM-sparse}
\end{figure}

The PDFs of the three aerodynamic coefficients considered in this study are displayed on \fref{fg:PDFCD-tensor} through \fref{fg:PDFCM-tensor} and \fref{fg:PDFCD-sparse} through \fref{fg:PDFCM-sparse} using the $10$--th level product rule and the $6$--th level sparse rule, respectively. They were estimated from $N_s=100,000$ evaluations of the gPC surrogates $\surr{\model}_\torder^\nquad$ and smoothing out the resulting histograms by a normal kernel density function~\cite{WAN95}. The means are shown on the plots with vertical blue lines.

\begin{figure}
\centering{\includegraphics[width=8.4cm]{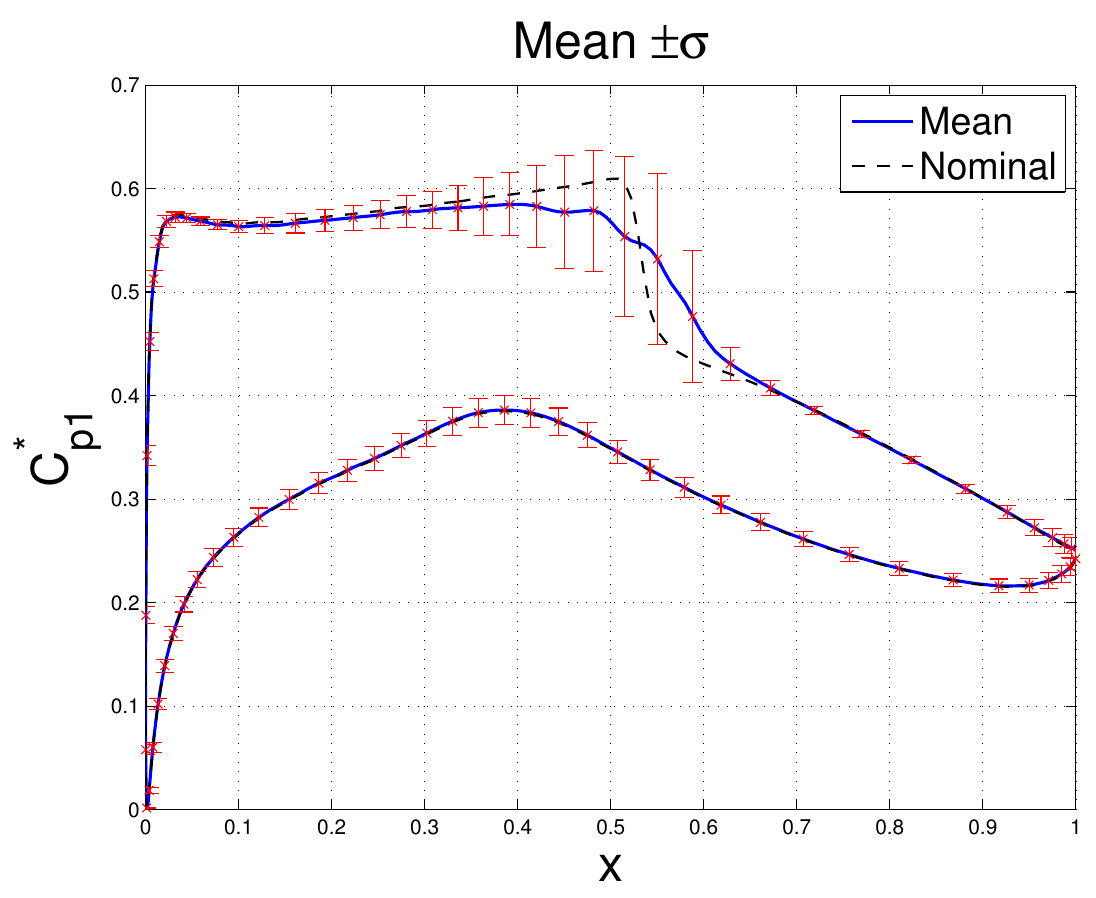}}
\caption{Total pressure coefficient computed by the $10$--th level product rule with $\nquad=10^3$.}\label{fg:Cp-totalT}
\end{figure}

\begin{figure}
\centering{\includegraphics[width=8.4cm]{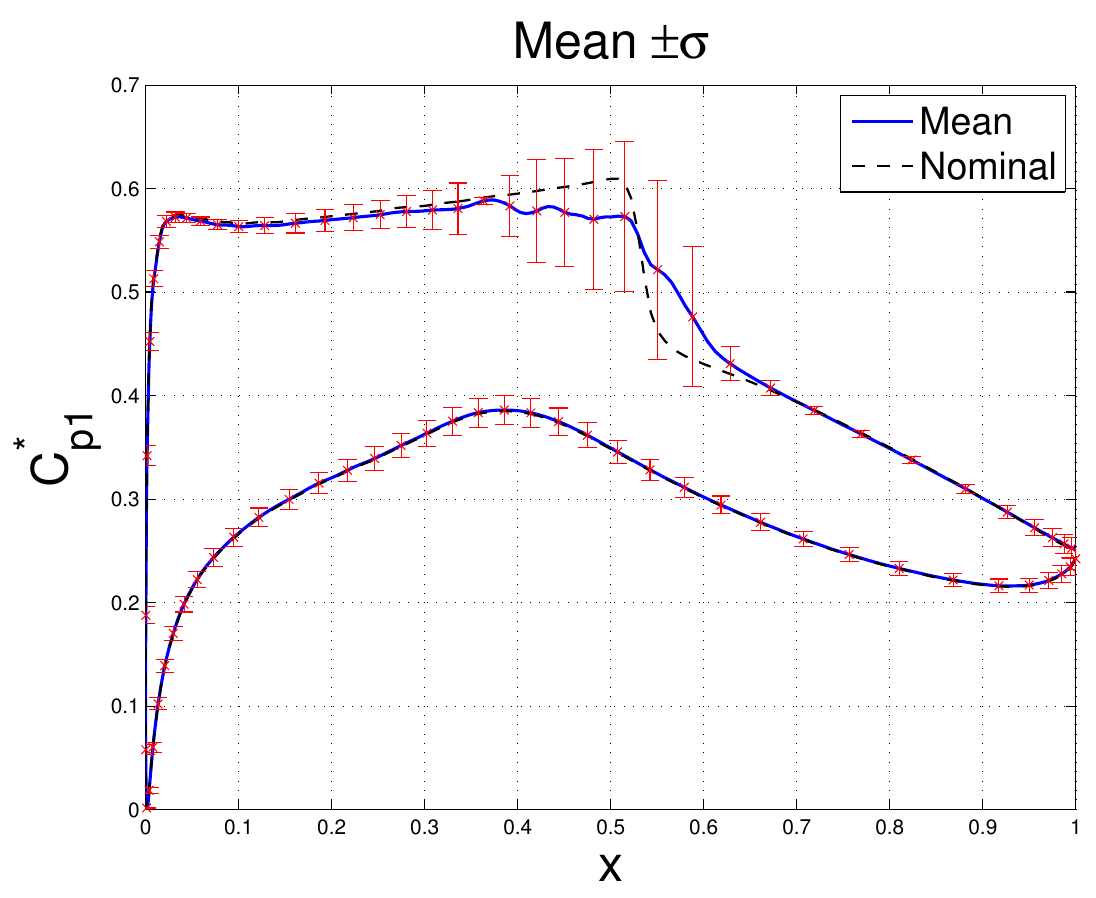}}
\caption{Total pressure coefficient computed by the $6$--th level sparse rule with $\nquad=201$.}\label{fg:Cp-totalS}
\end{figure}

Finally, the mean total pressure coefficients $C_{p1}^*$ along the profile are displayed on \fref{fg:Cp-totalT} using the $10$--th level product rule and on \fref{fg:Cp-totalS} using the $6$--th level sparse rule. Error bars at $\pm\sigma$, where $\sigma$ is the standard deviation of the $C_{p1}^*$'s, are also shown, together with the nominal total pressure coefficient (dotted line) obtained from a computation with the nominal parameters $\nominal{M}_\infty=0.729$, $\nominal{\alpha}=2.31^\text{o}$, $\nominal{Re}=6.50\cdot10^6$, and the nominal thickness-to-chord ratio. We observe an unexpected drop of the standard deviation at $\xgj\simeq0.36$ locally at the suction side when the $6$--th level sparse rule is used. This may be related to the negativeness of some weights with sparse rules, however we have not been able to find a more detailed explanation of this probable anomaly so far.  

\section{Non-adapted sparse reconstruction by $\ell_1$--minimization}\label{sec:CS}

The results of the previous section, especially the main-effect and joint sensitivity indices, suggest that only low-order interactions exist between the variable parameters for their effect on the aerodynamic coefficients of interest. This indicates that among the gPC coefficients to be computed for the construction of their polynomial surrogates $\smash{\surr{\model}_\torder}$, the ones pertaining to the one-dimensional polynomials (depending on a single variable parameter) dominate the others. Hence the vector $\smash{\modelg}=\smash{(\model_0,\model_1,\dots\model_\cardinal)^\itr\in\Rset^{\cardinal+1}}$ of gPC coefficients of $\smash{\surr{\model}_\torder}$ has many negligible components, so that it is \emph{compressible} in the terminology of the theory of compressed sensing~\cite{CAN06,DON06,CAN08}. In this setting it is argued that only a number of samples proportional to the compressed size, rather than the uncompressed size, of the sought signal is needed in order to reconstruct it. This observation challenges the conventional approaches to sampling or imaging according to Shannon's theorem, which states that the sampling rate (the Nyquist rate) must be at least twice the maximum frequency present in the signal. Compressed sensing, or compressive sampling (CS), asserts that one can recover some signals from far fewer samples or measurements than traditionally used in the widespread signal acquisition techniques. CS relies on two principles to make this possible:
\begin{enumerate}
\item Sparsity, which express the fact that many signals may have a concise representation once they are expressed in a proper basis $\mbase$;
\item Incoherence, which express the fact that signals having a sparse representation in a given basis $\mbase$ are actually spread out in the domain in which they are acquired. Or in other words, the sensing functions $\msensing$ used to probe the signal have a dense representation in the basis $\mbase$.
\end{enumerate}
The reconstruction procedure consists in correlating the signal with a small number of predefined sensing functions (for example, sinusoids if one aims at computing a Fourier transform) which are incoherent with the basis in which the signal is sparse. It is non adapted because it identifies the sparsity pattern, that is the order (location) of the negligible components of the signal in its sparsifying basis, and the leading components at the same time. The procedure can therefore efficiently capture the relevant information of a sparse signal without trying to comprehend that signal~\cite{CAN08}. This is clearly a much desirable feature for practical industrial applications.

\subsection{Theoretical background}

We basically follow the introductory paper of Cand\`es \& Wakin~\cite{CAN08} in this short presentation of CS. The sparse signal to be reconstructed is the polynomial surrogate model $\surr{\model}_\torder$ of total order $\torder$, for which information is obtained by recording $\nquad$ values of the model $\model$:
\begin{equation}\label{eq:sensing}
\QoI_\ell=\dual{\model,\vsensing_\ell}\,,\quad1\leq\ell\leq\nquad\,.
\end{equation}
Typically $\vsensing_\ell$ is a sinusoid and $\QoI_\ell$ is a Fourier coefficient if a Fourier transform is applied, or a Dirac function $\vsensing_\ell(\xg)=\delta(\xg-\xig_\ell)$ if the model $\model$ is evaluated at some sampling point $\xig_\ell$: $\smash{\QoI_\ell=\model(\xig_\ell)}$. This latter sensing procedure is the one considered in this work. Letting $\msensing$ be the $\nquad\times(\cardinal+1)$ sensing matrix of which rows are the vectors $\smash{\cjg{\vsensing}_1,\cjg{\vsensing}_2,\dots\cjg{\vsensing}_\nquad}$ ($\cjg{\boldsymbol a}$ is the conjugate transpose of ${\boldsymbol a}$), the process of recovering a discretized version $\smash{\modelg\in\Rset^{P+1}}$ of the model $\model$ from the observation of $\nquad$ outputs $\vQoI=\smash{(\QoI_1,\QoI_2,\dots\QoI_\nquad)^\itr}$ reads:
\begin{displaymath}
\vQoI=\msensing\modelg\,.
\end{displaymath}
This problem is generally ill-posed whenever $\nquad<\cardinal+1$, but CS theory tells us that a unique solution may be obtained if the vector $\smash{\modelg}$ is sparse. This is actually the case once the model $\model$ is expanded on the orthonormal basis constituted by the polynomial chaos pertaining to the randomly variable parameters introduced in section \ref{sec:BC02}. This property is observed \emph{a posteriori} for the present problem from the results expounded in the foregoing section \ref{sec:gPC-BC02}. Incidentally, it should be noted that sparsity can be proved \emph{a priori} for some parametric, possibly non-linear elliptic and parabolic partial differential equations in a general framework; see Chkifa \emph{et al}.~\cite{CHK14} and references therein. However, the extension of these analyses to the present RANS system supplemented with a turbulence closure model, not to say the Navier-Stokes equations, does not seem actually feasible.

The discretized version of the model $\model$ in terms of the polynomial surrogate $\surr{\model}_\torder$ reads $\smash{\model\approx\surr{\model}_\torder}=\smash{\sum_{j=0}^\cardinal\model_j\vbase_j}$ in its sparse representation on the basis $\smash{\mbase\equiv\basis^\torder}$. Applying the sensing procedure (\ref{eq:sensing}) to the polynomial surrogate yields:
\begin{equation}\label{eq:sensing_sparse}
\vQoI=\mmeasure\modelg^\nquad\,,
\end{equation}
where $\smash{\modelg^\nquad}=\smash{(\model_0^\nquad,\model_1^\nquad,\dots\model_\cardinal^\nquad)^\itr\in\Rset^{\cardinal+1}}$ is the sparse vector of the gPC coefficients in the basis $\mbase$ to be computed, and $\mmeasure$ is the so-called $n\times(\cardinal+1)$ measurement matrix given by $\smash{[\mmeasure]_{\ell j}}=\dual{\vbase_j,\vsensing_\ell}$. Using $\vsensing_\ell(\xg)=\delta(\xg-\xig_\ell)$ as done here, it is a Vandermonde-type matrix with $[\mmeasure]_{\ell j}=\vbase_j(\xig_\ell)$. Again, \eref{eq:sensing_sparse} is ill-posed whenever $\nquad\ll\cardinal$, unless the sparsity of the sought solution is accounted for. Regularized versions of \eref{eq:sensing_sparse} exists for this case, which in turn ensure its well-posedness. Since the polynomial chaos expansion truncated at the total order $\torder$ is not complete for the exact representation of $\model$, a truncation error has also to be accounted for in the solution process. Together with the sparsity of the gPC coefficients, this can be accommodated by reformulating \eref{eq:sensing_sparse} as the following $\ell_1$--minimization problem, known as Basis Pursuit Denoising (BPDN) \cite{CHE98}:
\begin{equation}\label{eq:BPDN}
\modelg^\nquad\approx\modelg^\star=\arg\min_{{\boldsymbol h}\in\Rset^{\cardinal+1}}\{\|\mweight\hg\|_1;\;\|\mmeasure\hg-\vQoI\|_2\leq\heps\}\,,\hfill\tag*{$(P_{1,\varepsilon})$}
\end{equation}
for some tolerance $0\leq\heps\ll 1$ on the polynomial chaos truncation. The norms above are defined by $\smash{\|\hg\|_m}=\smash{(\sum_{j=0}^\cardinal\hgj_j^m)^{\frac{1}{m}}}$, and $\mweight$ is some diagonal weighting matrix. Its role is to prevent the algorithm from biasing toward the non-negligible entries of $\smash{\modelg^\nquad}$ of which associated columns in $\mmeasure$ have large norms~\cite{DOO11,MAT12,PEN14}. Now the strategy for our present study is to solve \ref{eq:BPDN} with $\nquad$ runs of the physical model $\model$ significantly lower than the number of coefficients to be identified. CS shows that it is achievable provided that the target $\modelg^\nquad$ is actually sparse, or nearly sparse (compressive), and some constraints on the measurement matrix are fulfilled.

As already stated above, a key requirement for the successful recovery of a sparse signal is incoherence between the sensing basis $\msensing$ and the representation basis $\mbase$. It is quantified by the following mutual coherence $0<\mu(\msensing,\mbase)\leq 1$:
\begin{equation}\label{eq:coherence}
\mu(\msensing,\mbase)=\max_{\tiny\begin{array}{c}1\leq j,k\leq\cardinal+1 \\ j\neq k\end{array}}\frac{|\vmeasure_j^\itr\vmeasure_k|}{\|\vmeasure_j\|_2\|\vmeasure_k\|_2}\,,
\end{equation}
where $\vmeasure_j$ stands for the $j$--th column of $\mmeasure$. It is a measure of how close to orthogonality the measurement matrix is. Based on this coherency measure, the following theorem from Cand\`es \& Romberg~\cite{CAN07} asserts that if $\surr{\model}_\torder$ is sufficiently sparse in $\mbase$, the recovery of its gPC coefficients by $\ell_1$--minimization is exact.
\begin{mytheorem}\label{th:th1}
Assume that $\surr{\model}_\torder$ is $S$--sparse on the gPC basis $\mbase$, that is it has at most $S$ non-zero entries. Then if $\nquad$ sampling points $\smash{\{\xig_\ell\}_{\ell=1}^\nquad}$ are selected at random to form the measurement matrix $\mmeasure$, and:
\begin{equation}\label{eq:th1}
\nquad\geq C\cdot\mu(\msensing,\mbase)\cdot S\cdot\log\cardinal
\end{equation}
for some constant $C>0$, the solution of $(P_{1,0})$ is exact with "overwhelming" (sic) probability.
\end{mytheorem}
More precise results with structured random measurement matrices are given by \emph{e.g.} Rauhut \& Ward~\cite{RAU12}. It should be noted that the role of coherence in this result is transparent. The smaller the coherence is, the closer the measurement matrix is to a unitary matrix, and the fewer sampling points are needed. 

The previous Theorem \ref{th:th1} is however not entirely satisfactory from a practical point of view because (i) it does not allow for some truncation error, or noisy/imprecise measurements; (ii) it does not deal with approximately sparse (compressive) signals, for which a large subset of entries are negligible rather than strictly zero. These shortcomings may be alleviated simultaneously as established by Cand\`es \emph{et al.}~\cite{CAN06}. To achieve this, a constraint on the measurement matrix $\mmeasure$ needs be added to gain robustness in CS, the so-called restricted isometry property (RIP, also quoted as uniform uncertainty principle). For each integer $S\in\Nset^*$, the isometry constant $\delta_S$ of $\mmeasure$ is defined as the smallest number such that:
\begin{displaymath}
(1-\delta_S)\|\hg_S\|_2^2\leq\|\mmeasure\hg_S\|_2^2\leq(1+\delta_S)\|\hg_S\|_2^2
\end{displaymath}
for all $S$-sparse vectors $\hg_S\in\QoIset_S:=\{\hg\in\Rset^{\cardinal+1};\;\|\hg\|_0\leq S\}$. Then $\mmeasure$ is said to satisfy the RIP of order $S$ if, say, $\delta_S$ is not too close to $1$. This property amounts to saying that all $S$--column submatrices of $\mmeasure$ are numerically well-conditioned, or $S$ columns $\smash{\vmeasure_{j_1},\vmeasure_{j_2}\dots\vmeasure_{j_S}}$ selected arbitrarily in $\mmeasure$ are nearly orthogonal. The following theorem by Cand\`es \emph{et al.}~\cite{CAN06,CAN08} then states that \ref{eq:BPDN} can be solved efficiently:
\begin{mytheorem}
Assume $\delta_{2S}<\sqrt{2}-1$. Then the solution $\modelg^\star$ to \ref{eq:BPDN} satisfies:
\begin{displaymath}
\|\modelg^\star-\modelg^\nquad\|_2\leq C_0\frac{\|\modelg_S^\nquad-\modelg^\nquad\|_1}{\sqrt{S}}+C_1\heps
\end{displaymath}
for some $C_0,C_1>0$. Here $\modelg_S^\nquad$ is $\modelg^\nquad$ with all but the $S$ largest entries set to zero.
\end{mytheorem}
This result calls for several comments. First, it is more general than Theorem \ref{th:th1} since, if the signal is exactly $S$--sparse, $\modelg^\nquad=\modelg_S^\nquad$ and the reconstruction is exact whenever $\heps=0$ (noiseless case). Second, it deals with all signals, not the $S$--sparse ones solely. Third, it is deterministic and does not involve any probability. Lastly, the bound $\sqrt{2}-1$ on $\delta_{2S}$ is the one originally proposed by Cand\`es \& Wakin~\cite{CAN08} but it can be improved as proposed by \emph{e.g.} Mo \& Li~\cite{MO11}; such improvements are an active field of research at present.

\subsection{Application to the two-dimensional RAE 2822 transonic airfoil}

We now apply the foregoing CS procedure to the non-adaptive computation of the gPC coefficients $\smash{\modelg^\nquad}$ of the surrogate models $\smash{\surr{\model}_\torder}$ for the aerodynamic coefficients $\smash{C_D}$, $\smash{C_L}$ and $\smash{C_M}$. We use $\nquad=80$ sampling points drawn at random following $\beta_{\mathrm I}(4,4)$ PDFs as defined in section \ref{sec:BC02}. This sampling set is displayed on \fref{fg:DOE-CS} below. The primary reason why we have chosen this sampling size is for its ease of use with the multithreading setup of our CFD software \emph{elsA}~\cite{CAM13,elsa}. However, the mutual coherence for the present sampling set and representation basis is $\smash{\mu(\msensing,\mbase)}\simeq0.93$ and the sparsity of the polynomial surrogates is observed to be $S\simeq10$ from the results of the section~\ref{sec:gPC-BC02}. Thus \eref{eq:th1} yields $\nquad\gtrsim50$. A common observation is that $\nquad\gtrsim 4S\simeq40$ is usually enough for a successful recovery.

\begin{figure}
\centering{\includegraphics[width=7.4cm]{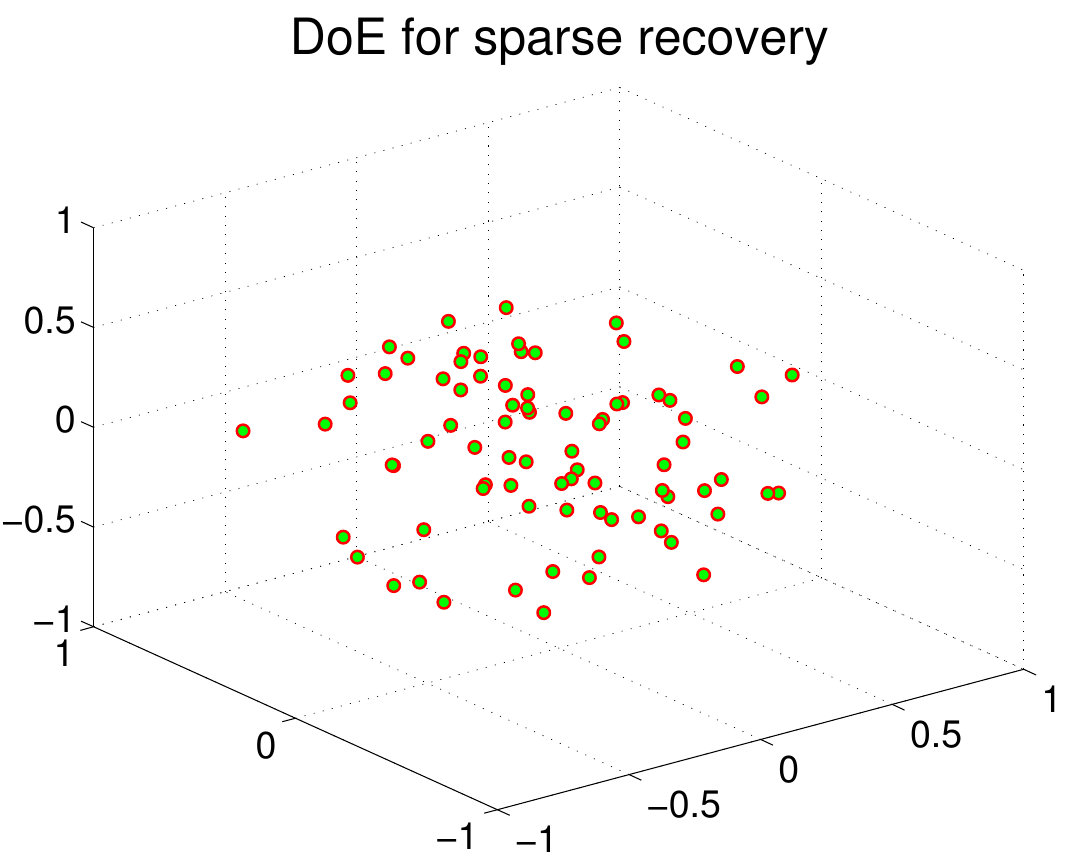}}
\caption{Sampling points used to non-adaptively compute the gPC coefficients of the polynomial surrogates of $\smash{C_D}$, $\smash{C_L}$ and $\smash{C_M}$ by $\ell_1$--minimization.}\label{fg:DOE-CS}
\end{figure}

We subsequently apply BPDN \ref{eq:BPDN} to compute $\smash{\modelg^\nquad}$. For that purpose we use the Spectral Projected Gradient Algorithm (SPGL1) developed by van den Berg \& Friedlander~\cite{VDB08} and implemented in the Matlab package \texttt{SPGL1}~\cite{SPGL1} to solve this $\ell_1$--minimization problem. The tolerance was fixed at $\heps=10^{-5}$ and we were able to find a solution for all surrogates with this \emph{a priori} choice without resorting to cross-validation, for example. Further investigations should be carried on on this topic, though. It should also be noted that no particular sampling strategy, such as stratification, low-discrepancy series, or preconditioning, has been applied at this stage to select the sampling set. Moreover, the weighting matrix $\mweight$ was selected as the identity matrix. Alternative sampling and weighting strategies are outlined in several recent works\cite{DOO11,MAT12,PEN14,RAU12,YAN12,YUA13,HAM15}.

\begin{figure}
\centering{\includegraphics[width=7.4cm]{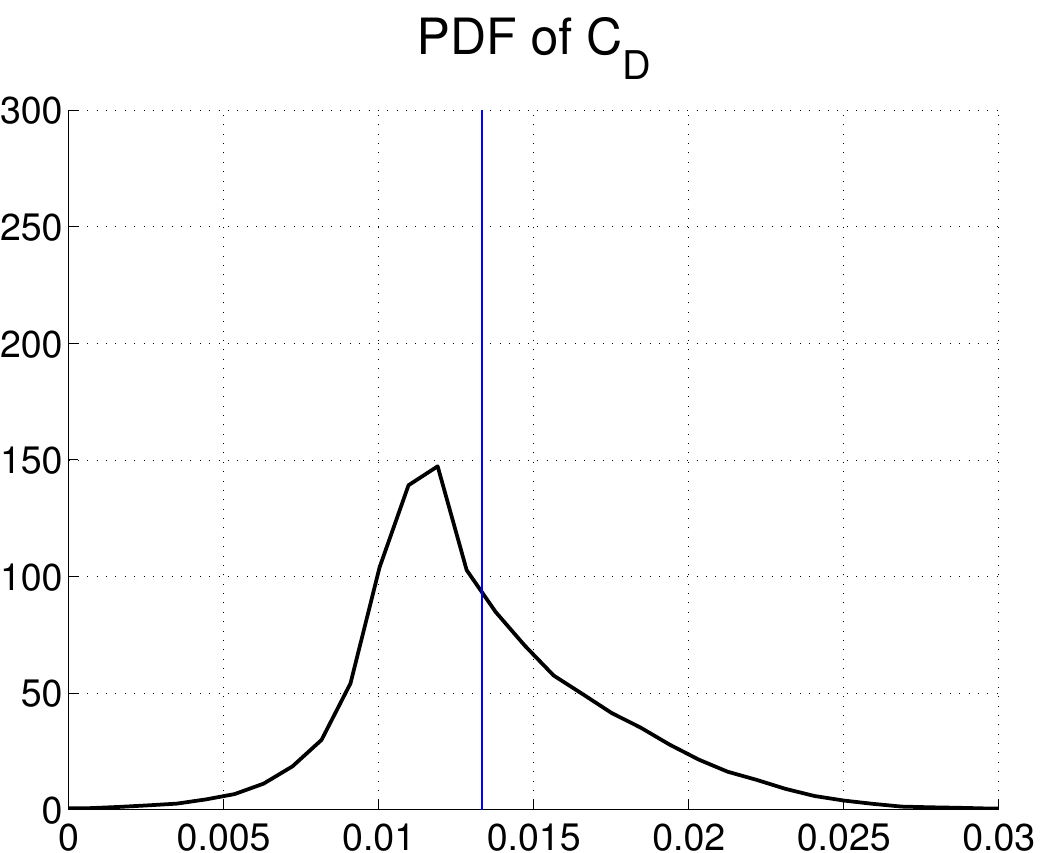}}
\caption{PDF of the drag coefficient $C_D$ computed by $\ell_1$--minimization with $\nquad=80$ random sampling points.}\label{fg:PDFCD-CS}
\end{figure}

\begin{figure}
\centering{\includegraphics[width=7.4cm]{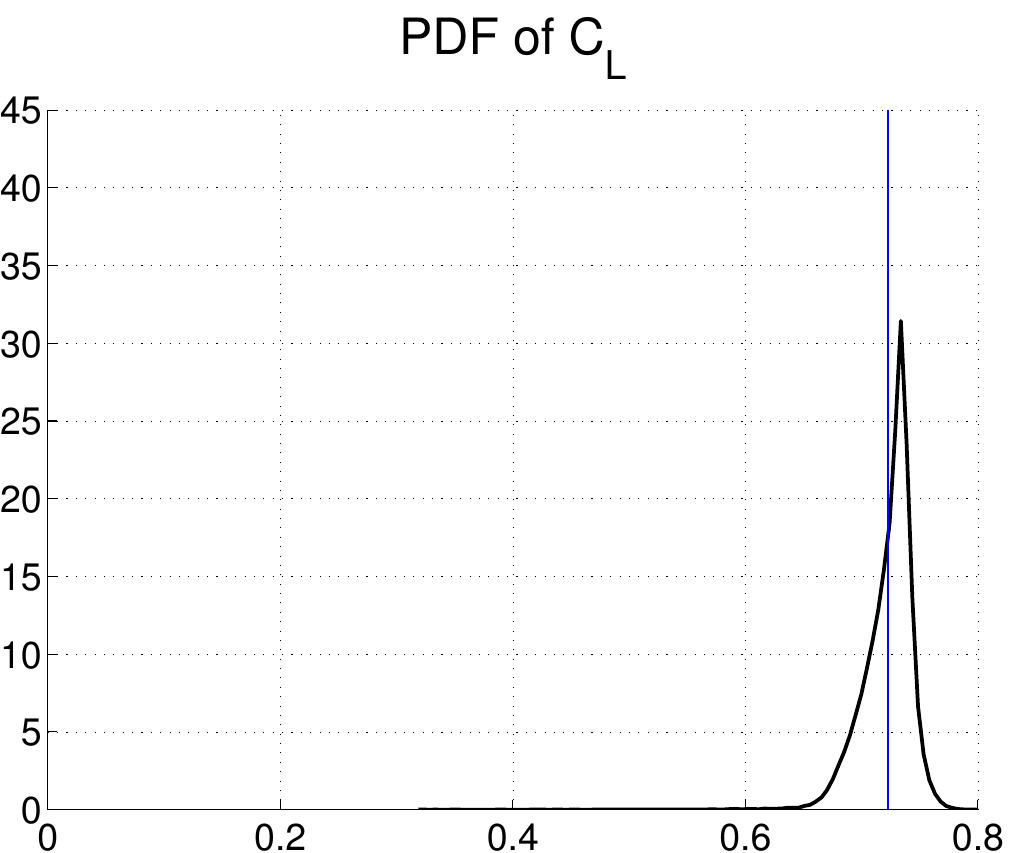}}
\caption{PDF of the lift coefficient $C_L$ computed by $\ell_1$--minimization with $\nquad=80$ random sampling points.}\label{fg:PDFCL-CS}
\end{figure}

\begin{figure}
\centering{\includegraphics[width=7.4cm]{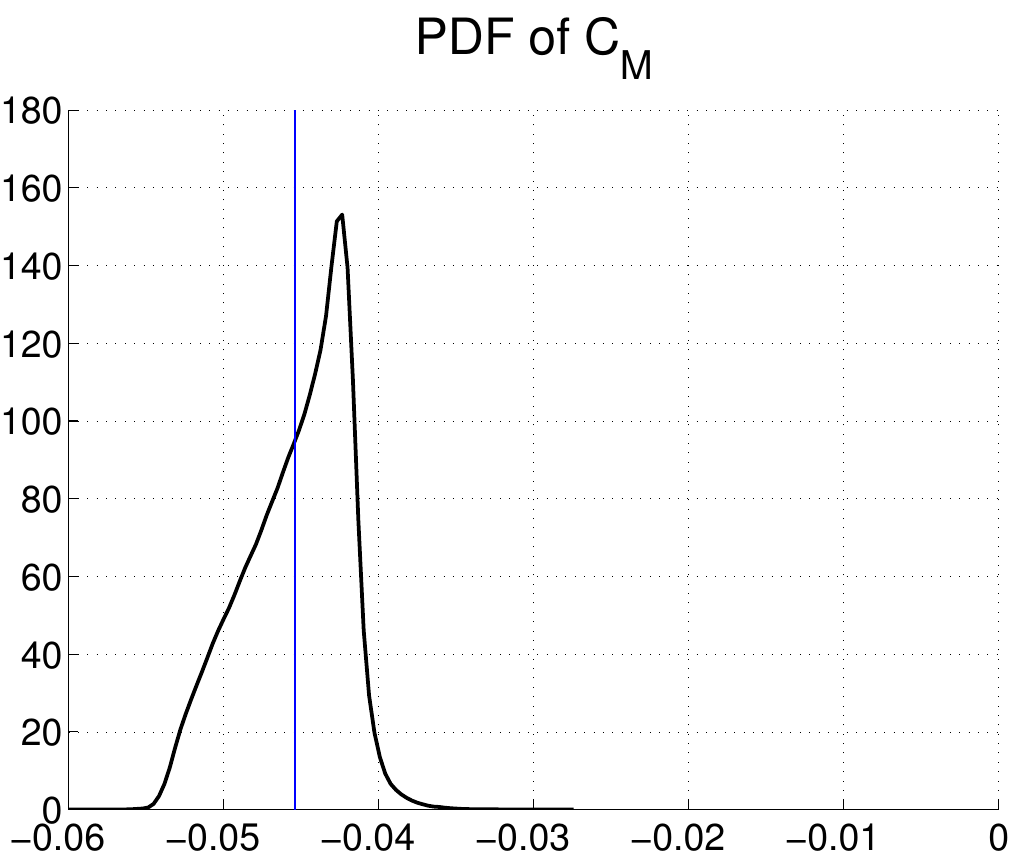}}
\caption{PDF of the pitching moment coefficient $C_M$ computed by $\ell_1$--minimization with $\nquad=80$ random sampling points.}\label{fg:PDFCM-CS}
\end{figure}

\begin{figure}
\centering{\includegraphics[width=7.4cm]{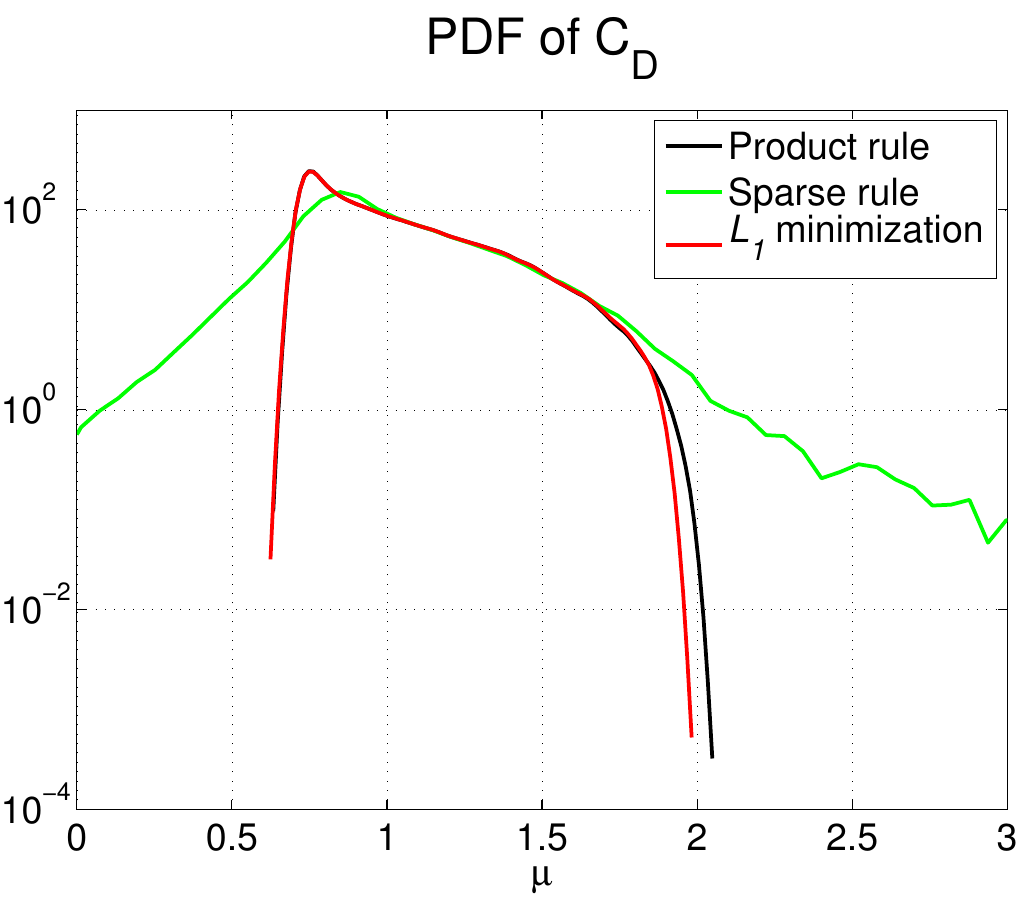}}
\caption{Comparison of the PDFs of the drag coefficient $C_D$ computed by the $10$--th level product rule ($\nquad=1,000$, black curve), the $6$--th level sparse rule ($\nquad=201$, green curve), and $\ell_1$--minimization ($\nquad=80$, red curve).}\label{fg:PDFCD-all}
\end{figure}

\begin{figure}
\centering{\includegraphics[width=7.4cm]{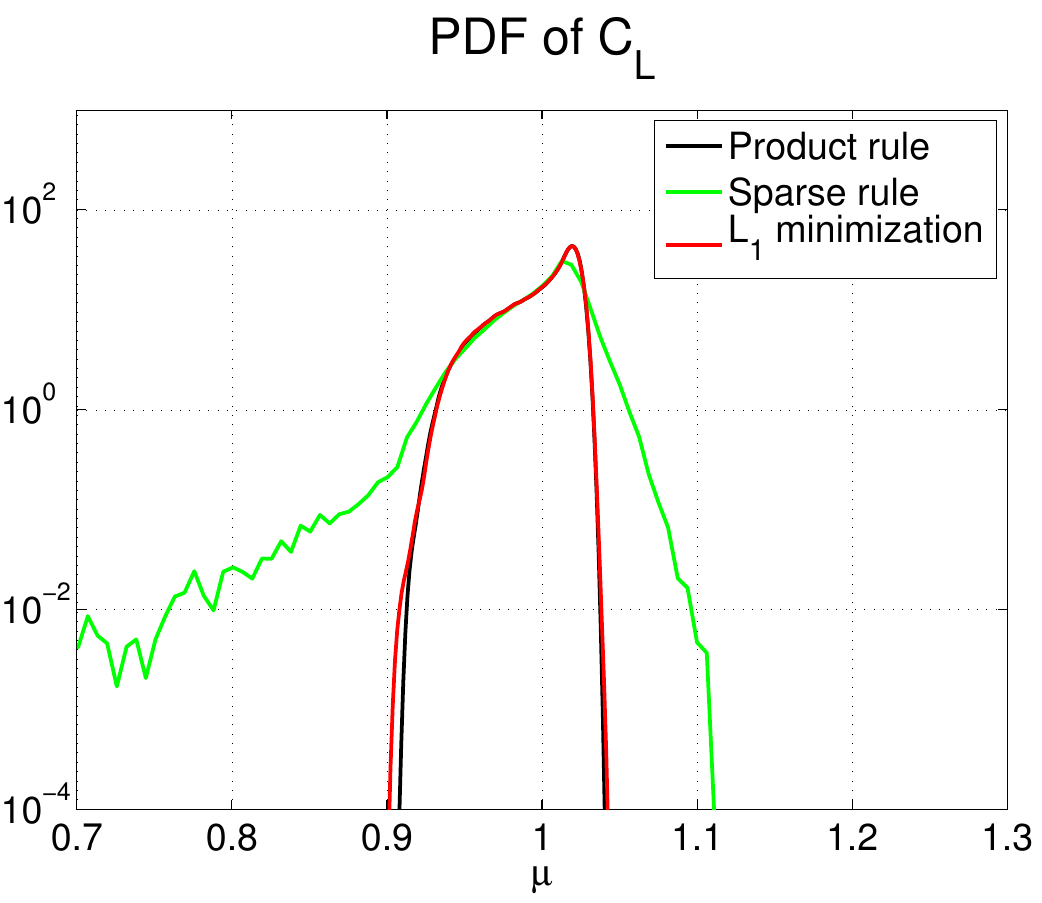}}
\caption{Comparison of the PDFs of the lift coefficient $C_L$ computed by the $10$--th level product rule ($\nquad=1,000$, black curve), the $6$--th level sparse rule ($\nquad=201$, green curve), and $\ell_1$--minimization ($\nquad=80$, red curve).}\label{fg:PDFCL-all}
\end{figure}

\begin{figure}
\centering{\includegraphics[width=7.4cm]{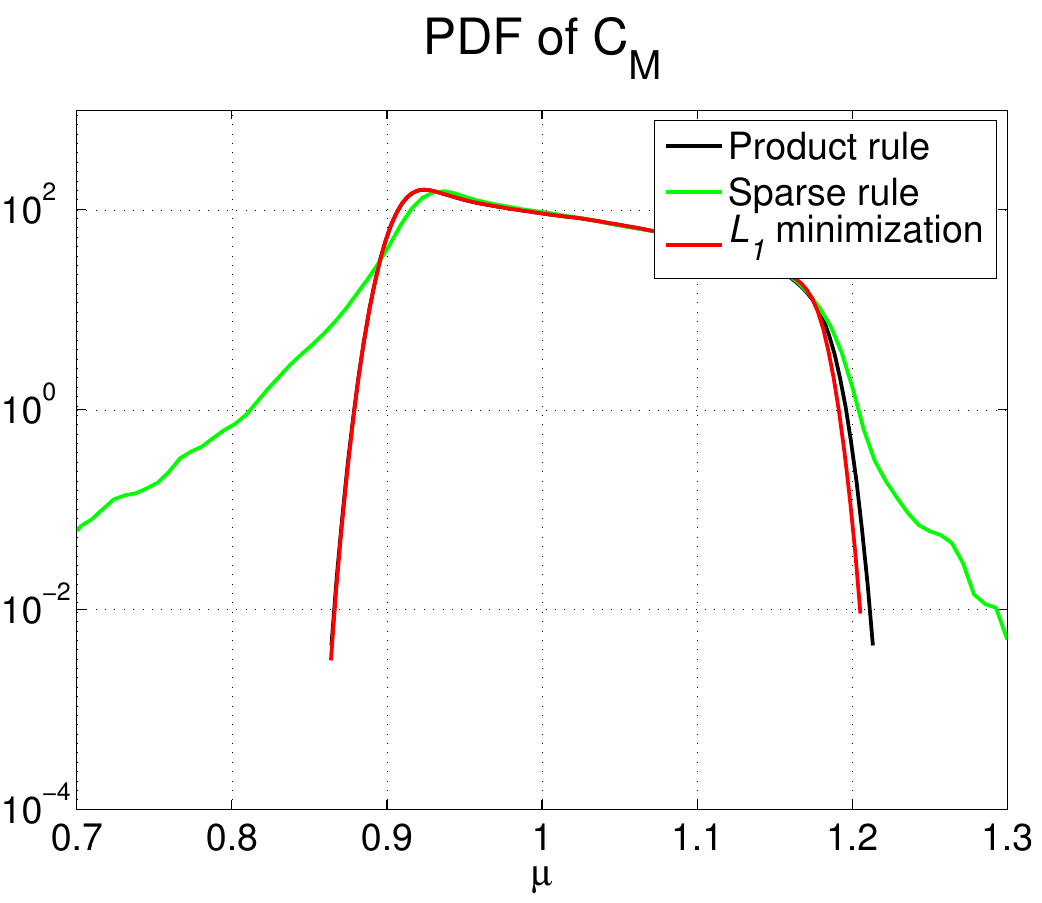}}
\caption{Comparison of the PDFs of the pitching moment coefficient $C_M$ computed by the $10$--th level product rule ($\nquad=1,000$, black curve), the $6$--th level sparse rule ($\nquad=201$, green curve), and $\ell_1$--minimization ($\nquad=80$, red curve).}\label{fg:PDFCM-all}
\end{figure}

The mean and standard deviation of the drag, lift, and pitching moment coefficients $C_D$, $C_L$, and $C_M$, respectively, are gathered in the Table \ref{tb:CS} below. The main-effect sensitivity indices are gathered in Table \ref{tb:Soboli-CS}, while the joint sensitivity indices are gathered in Table \ref{tb:Sobolij-CS}. These results are very close to the ones obtained by the $10$--th level product rule. The PDFs of the aerodynamic coefficients considered here are displayed on \fref{fg:PDFCD-CS} through \fref{fg:PDFCM-CS}. As for \fref{fg:PDFCD-tensor} through \fref{fg:PDFCM-sparse}, they were estimated from $N_s=100,000$ evaluations of the gPC surrogates $\surr{\model}_\torder^\nquad$ and smoothing out the resulting histograms by a normal kernel density function~\cite{WAN95}. The means are again shown on the plots with vertical blue lines. We finally compare on \fref{fg:PDFCD-all} through \fref{fg:PDFCM-all} (in $\log$ scale) the PDFs computed by the three approaches considered in this work. We observe that the $10$--th level product rule (with $\nquad=1,000$ structured sampling points) and $\ell_1$--minimization (with $\nquad=80$ randomly selected sampling points) yield comparable results, but of course at a much lower computational cost with this latter technique.

\begin{table}[h!]
\begin{center}
\begin{tabular}{|c||c|c|}
\hline
 & \makebox[3em]{$\mu$} &  \makebox[3em]{$\sigma$} \\
\hline\hline
$C_D$ & 133.33e-04 & 34.052e-04 \\
$C_L$ & 72.271e-02 & 1.6703e-02 \\
$C_M$ & -453.95e-04 & 32.180e-04 \\
\hline
\end{tabular}
\end{center}
\caption{Mean and standard deviation of the aerodynamic coefficients computed by $\ell_1$--minimization with $\nquad=80$ random sampling points.}\label{tb:CS}
\end{table}

\begin{table}[h!]
\begin{center}
\begin{tabular}{|c||c|c|c|c|}
\hline
& \makebox[3em]{$\xigj_1=h/c$} &  \makebox[3em]{$\xigj_2=M_\infty$} & \makebox[3em]{$\xigj_3=\alpha$} \\
\hline\hline
$C_D$ & 0.080e-01 &  9.893e-01 & 0.008e-01 \\
$C_L$ & 0.032e-01 &  9.549e-01 & 0.289e-01 \\
$C_M$ & 0.268e-01 &  9.722e-01 &  0.000e-01\\
\hline
\end{tabular}
\end{center}
\caption{Main-effect sensitivity indices of the aerodynamic coefficients computed by $\ell_1$--minimization with $\nquad=80$ random sampling points.}\label{tb:Soboli-CS}
\end{table}

\begin{table}[h!]
\begin{center}
\begin{tabular}{|c||c|c|c|c|}
\hline
& \makebox[3em]{$\xigj_2\xigj_3$} &  \makebox[3em]{$\xigj_1\xigj_3$} & \makebox[3em]{$\xigj_1\xigj_2$} \\
\hline\hline
$C_D$ & 0.022e-02 & 0.000e-02  & 0.166e-02 \\
$C_L$ & 0.031e-02 & 0.003e-02 & 1.265e-02 \\
$C_M$ & 0.007e-02 & 0.000e-02 & 0.093e-02 \\
\hline
\end{tabular}
\end{center}
\caption{Joint sensitivity indices of the aerodynamic coefficients computed by $\ell_1$--minimization with $\nquad=80$ random sampling points.}\label{tb:Sobolij-CS}
\end{table}

\section{Conclusions} \label{sec:conclusions}

In this paper we have addressed various methodologies with relevance to the construction of polynomial surrogates for parameterized complex processes as encountered in CFD problems. The present work has been more particularly focused on the development of dedicated sampling sets in the parameter space using either structured or unstructured grids. These techniques were illustrated with the example of an RAE 2822 airfoil in the transonic regime considering variable geometrical (the thickness-to-chord ratio) and operational (the free-stream Mach number and angle of attack) parameters. 

Firstly, multi-dimensional sparse cubature rules based on one-dimensional Gauss-Jacobi rules have been used for uncertainty quantification of this two-dimensional aerodynamic computation. The quantities of interest are the usual drag, lift, and pitching moment coefficients for which polynomial surrogates are sought for using the aforementioned sampling sets as learning sets. More particularly, Gauss-Jacobi-Lobatto points have been considered since the probability density functions of the variable parameters have finite supports. Indeed, the engineering practice would typically include the boundary points of the parameter space in the learning sets.

Secondly, observing \emph{a posteriori} that the aerodynamic quantities of interest are sparse in that parametric space, when projected on the multi-dimensional orthogonal polynomials associated to the parameters probability density functions, an $\ell_1$--minimization procedure has been applied in the framework of the theory of compressed sensing. The latter allows to recover the expansion coefficients of the quantities of interest at a much lower computational cost than the sparse grids addressed in the first approach. Unstructured sampling points are needed in this process, selecting them randomly in the parameter space. Their number is typically less than the dimension of the polynomial space where the surrogates are sought for, and thus typically much less than the number of points of the multi-dimensional sparse rules that may be used for a given level of accuracy. The $\ell_1$--minimization procedure is non-adaptive in the sense that it identifies both the amplitude of the leading expansion coefficients and their order. It thus constitutes a promising direction for future developments in practical applications for more complex geometries and flows, where adaptive strategies within the parametric space, weighted minimization, or preconditioned sampling sets may be needed.

%

\section*{Acknowledgments}

This work has been supported by the European Union's Seventh Framework Programme for research, technological development and demonstration under grant agreement \#ACP3-GA-2013-605036 (UMRIDA Project \href{http://www.umrida.eu}{\texttt{www.umrida.eu}}).

\end{document}